\documentclass[bibyear]{aa} 

\usepackage{graphicx}
\usepackage{txfonts}
\begin{document}

   \title{Stellar content of extremely red
quiescent galaxies at $z>2$}

   \subtitle{}

   \author{M. L\'opez-Corredoira\inst{1,2}, A. Vazdekis\inst{1,2},
 C. M. Guti\'errez\inst{1,2} \and N. Castro-Rodr\'\i guez\inst{3}}

   \institute{$^1$ Instituto de Astrof\'\i sica de Canarias, 
E-38205 La Laguna, Tenerife, Spain\\
$^2$ Departamento de Astrof\'\i sica, Universidad de La Laguna,
E-38206 La Laguna, Tenerife, Spain\\
$^3$ GRANTECAN S. A., E-38712, Bre\~na Baja, La Palma, Spain}

   \date{Received xxxx; accepted xxxx}

 
  \abstract
  {A set of 20 extremely red galaxies at $2.5\le z_{\rm phot.}\le 3.8$ with photometric features of old passive-evolving galaxies without dust, with stellar masses of $\sim 10^{11}$ M$_\odot $, have colors that could be related to passive-evolving galaxies with mean ages larger than 1 Gyr. This suggests they have been formed, on average, when the Universe was very young ($<1$ Gyr).}
   {We provide new estimates for the stellar content of these 20 galaxies, with a deeper analysis for two of them that includes spectroscopy.}
   {We obtained, with the GRANTECAN-10.4\,m, ultraviolet rest-frame spectra of two galaxies and analyzed them together with photometric data. The remaining 18 galaxies are analyzed only with photometry. We fit the data with models of a single-burst stellar population (SSP), combinations of two SSPs, as well as with extended star formation.}
   {Fits based on one SSP do not provide consistent results for the blue and red wavelengths. Moreover, the absence in the spectra
of a break at $\sim 2\times 10^3$ \AA \ indicates that a rather young component is necessary. 
Using two SSPs we can match the photometric and spectroscopic data, with the
bulk of the stellar population being very old (several Gyr)
and the remaining contribution ($<5$\% of stellar mass fraction) from a young, likely residual star formation component with age $\lesssim 0.1$ Gyr.
Exponentially decaying extended star formation 
($\tau $) models improve slightly the fits with respect to the single burst model, but they are considerably worse than the two SSP based fits, further supporting the residual star formation scenario.}
   {The fact that one SSP cannot match these
early-type galaxies highlights the limitations for the use of age estimators based on single lines or breaks, such as the Balmer break used in cosmic chronometers, thus questioning this approach for cosmological purposes.}


\titlerunning{Red quiescent galaxies $z>2$}
\authorrunning{L\'opez-Corredoira et al.}

   \maketitle
%

\section{Introduction}

Ages of quiescent galaxies have been derived by fitting photometric data with stellar
population synthesis models since the pioneering works for low redshift galaxies
of, for example, Tinsley (1980), and at redshifts larger than one since more than a
decade ago.
Early-type massive extremely red objects with $z\sim 1.4$
(e.g., Longhetti et al. 2005, Trujillo et al. 2006) have been found to be old. 
In fact, a galaxy at $z=1.22$  has an average age of its
stellar population of around 5.0 Gyr (Longhetti et al. 2005), 
within the limit of the age of the
Universe at this redshift (5.3 Gyr).
More distant ($z>2$) red galaxies were analyzed (Labb\'e et al. 2005)
with observations that provide rest-frame ultraviolet to infrared photometry (0.3-8.0 $\mu $m). These observations found that three out of 14 red galaxies were indeed old galaxies with mean luminosity-weighted ages of 
2.6 Gyr at $z=2.7$, 3.5 Gyr at $z=2.3$,
and 3.5 Gyr at $z=2.3$ respectively, representative of
the average age of the stellar populations. 
The age of the Universe in standard models at both redshifts of $2.7$
and $2.3$ is 2.5 Gyr and 3.0 Gyr respectively.
These are mean luminosity-weighted ages assuming a single
stellar population.
Toft et al. (2005) fit synthesis models to spectra of red compact
galaxies with ages 5.5, 3.5, and 1.7 Gyr for redshifts 1.2, 1.9, and 3.4 (ages
of the Universe respectively: 5.4, 3.6, 2.0 Gyr), within the limit.
Galaxies at very high redshift ($z\sim 4$)
were analyzed (Rodighiero et al. 2007) to find, with optical, near-infrared 
and mid-infrared surveys, an important ratio of old 
($\gtrsim 1$ Gyr) and massive ($\sim 10^{11}$ M$_\odot $) galaxies.
With similar techniques,
evidence was presented for 11 massive and evolved (0.2-1.0 Gyr)
galaxies at redshifts $4.9\le z_{\rm phot.} \le 6.5$ (Wiklind et al.
2008). Therefore, it is clear that there are
bright massive galaxies at $z>2.5$ (e.g., Alcalde Pampliega et al. 2015). 
It is clear from this information that the formation of very massive elliptical
galaxies takes place at very high redshifts. These tight limits to the age of the galaxies can be in principle used to constrain the age of the Universe at different redshifts, and consequently to constrain the cosmological models (e.g., Simon et al. 2005), although there are uncertainties such as those
from the photometric redshifts or the fact that the age estimation should be understood in relative terms, especially when galaxies are
very old (Age Zero Point issue; Vazdekis et al. 2001, Schiavon et al. 2002).
The existence of very massive galaxies almost as old as the Universe is in tension with semianalytic $\Lambda $CDM models, which claim that these galaxies were assembled much later (Guo et al. 2011). This is what was called the impossible early galaxy problem, 
by which observations find, by several orders of magnitude, more very massive haloes
at very high redshift than predicted, implying that these massive galaxies formed impossibly early (Steinhardt et al. 2016).
This is along the line suggested by the ``downsizing'' scenario of galaxy formation in which most massive galaxies assembled important amounts of their stellar content rapidly (in 1-2 Gyr) beyond $z\sim 3$ in very intense star formation events (P\'erez-Gonz\'alez et al. 2008).

Another example of the exploration of the age of quiescent galaxies is
given in the paper by Castro-Rodr\'\i guez \& L\'opez-Corredoira (2012, hereafter CL12). Among a sample in the XMM-LSS field of $\sim 60$ thousand galaxies with deep photometry in the visible (Subaru telescope), near infrared (UKIDSS-Ultra Deep Survey) and mid infrared (IRAC), with redshifts derived photometrically, CL12 searched for extremely red galaxies at $z_{\rm phot.}\ge 2.5$ with photometric features of old passively evolving galaxies and without dust (see further details in \S \ref{.high}). They prove to be massive galaxies (stellar masses of $\sim 10^{11}$ M$_\odot $) which were formed on average when the Universe was very young ($<1$ Gyr), assuming the standard cosmological parameters (CL12) and one single-burst stellar population (SSP).

The problem becomes more complex if we bear in mind that these elliptical galaxies that host an old, virtually single, stellar population (Renzini 2006)
may have a small residual component from young stellar populations, as observed in local galaxies (Vazdekis et al. 1997; Atlee et al. 2009). This is more relevant at high redshift because the optical range samples the ultraviolet (UV) at rest, and
this range is particularly sensitive to the hottest stars of the youngest stellar population components. Determining such contributions is relevant for constraining the latest generation of cosmological hydro-dynamical simulations like EAGLE, which predict some small level of recent star formation (Trayford et al. 2015). This is a key aspect of the work presented here, as opposed to many studies that often assume one SSP, which are biased toward
the youngest contributions for fitting their data, and therefore the maximum age of the oldest population of the galaxy may be underestimated. 
Deriving accurately the amount of young stellar populations
in early-type galaxies is of paramount relevance to their formation within
the $\Lambda $CDM model (e.g., Kaviraj et al. 2007) and particularly to
constrain whether such contributions are consistent with residual star formation
after an initial monolithic-like rapid formation phase (e.g., Oser et al. 2010),
or in an extended formation assuming a delayed exponentially declining star formation history (Dom\'\i nguez-S\'anchez et al. 2016). 
Moreover, the fact that one SSP may not match these
early-type galaxies highlights the limitations for the use of age estimators 
based on single lines or breaks, such as the Balmer break used in cosmic chronometers (Moresco et al. 2012, 2016), thus questioning this approach for cosmological purposes. 

Here, we aim to perform a more accurate analysis of the stellar populations for the objects of the CL12 sample, modeling the photometry and
spectra with one or two SSPs.
There are studies of spectra in passively evolving galaxies up to $z\approx 2$ (e.g., 
McCarthy et al. 2004; Cimatti et al. 2004, 2008; Mignoli et al. 2005; Toft et al. 2012;
Whitaker et al. 2013; Bedregal et al. 2013;
Geier et al. 2013; Belli et al. 2015; Mendel et al. 2015; Onodera et al. 2015; Lonoce et al.
2015). The challenge here is to study low resolution spectra of faint galaxies ($m_R=24-25$) dominated by absorption features at $z>2$. We will also use the same photometric data used by CL12 for all of their 20 galaxies, 
and we will constrain the age and redshift of the galaxies with better accuracy.
By means of very-low resolution spectra, we test further this result through a more accurate method for two of these galaxies, \#40812 and \#78891, with photometric redshifts determined by CL12 of $z=2.57$ and $z=2.71$ respectively. We will explore the spectral range from 5\,000 to 9\,000 \AA , which at rest-frame covers the range around 2,000 \AA. That UV range has an abrupt break (see Fig. 9 of Bruzual \& Charlot (2003, hereafter BC03)) that depends strongly on the age of the stellar population, so these spectra can be useful
for this purpose. The low resolution spectra in this range allow us to fit the best template with its corresponding age. The metallicity effect will also be considered, whose effect in the change of the continuum of the spectra is smaller at such young ages. 

Section \ref{.data} describes the photometric and spectroscopic data used in this paper. Section \ref{.models} describes two sets of stellar population
synthesis models used here. The results of the best fit ages and other parameters for the two selected galaxies are given in Sections \ref{.fitting}, \ref{.fit2}, and \ref{.fitother}. In  Section \ref{.others}, we do some similar analysis only with photometry for the remaining 18 galaxies of the list. Discussions and conclusions are presented in Sections \ref{.discus} and \ref{.conclus}. 

\section{Data}
\label{.data}

\subsection{High redshift sample}
\label{.high}

In CL12, the photometric selection was carried out with color-color diagrams similar to those of Pozzetti \& Mannucci (2000) or Fang et al. (2009) in the visible/near-infrared, but shifted towards longer wavelengths.
A negligible emission in 24 $\mu $m was also required to further reduce the chances of obtaining a starburst in the selection. Moreover, CL12 selected objects for which the photometrically derived redshifts
have been confirmed by different authors and different algorithms within $\frac{\Delta z}{(1+z)}<0.1$. In total, CL12 obtained a sample of 20 galaxies with $2.5\le z_{\rm phot.}\le 3.8$. From these galaxies, the rest $(B-V)$ color was determined through spectral energy distribution (SED) using templates of galaxies with the software Interrest v2.0 (Taylor et al. 2009; L\'opez-Corredoira 2010). Average colors $(B-V)$ between 0.7 and 0.8 (Vega calibration) were obtained. There is negligible extinction and reddening due to our Galaxy. From the comparison with a stellar population synthesis model and solving the degeneracy with metallicity by means of an iterative method which takes into account the mass-metallicity relationship (method developed by L\'opez-Corredoira 2010 using the models by Vazdekis et al. 1996), these colors could be related to ages above 1 Gyr of a stellar population in passive-evolving galaxies.
We use two sets of data for two galaxies from the CL12 sample, 
numbers \#40812 (R.A.=2h16m42.40s, $\delta =$-5$^\circ$ 8' 6.5'')
and \#78891 (R.A.=2h18m21.99s, $\delta =$-4$^\circ$ 46' 49.4''), corresponding to the number of sources in the Ultra Deep Survey XMM-LSS.

\subsection{Photometric data}
\label{.photdata}

For these galaxies, we use the same photometric data as CL12, with the filters B, V, R, i, z, J, H, K, [3.6$\mu $m], and [4.5$\mu $m] (only for \#40812).
We correct the photometry of Galactic extinction, with $A_R=0.047$ 
and $A_R=0.045$ (Schlafly \& Finkbeiner 2011) respectively for the galaxies \#40812 and \#78891.
Similar photometric analysis will be done for the remaining 18 galaxies of CL12
in \S \ref{.others}.

\subsection{Spectroscopic data}

Spectroscopic observations of the optical counterparts were made with the OSIRIS spectrograph on the 10.4 m GTC in La Palma (Canary islands, Spain) using its service observing mode on August 29 and 31 2014. Long-slit spectra of the R300R grism/binning 2x2 over the range 4800-10000 \AA \ were obtained, at the parallactic angle of each exposure, under 0.8-0.9'' seeing condition and a slit width of 1''. The total exposure time was 2670 s for galaxy \#40812, and 8010 s for galaxy \#78891. The spectral resolution with this configuration was 7.75 \AA/pixel; but, in order to gain signal/noise ratio (S/N), we binned the spectrum in bins of three and ten pixels respectively for both galaxies, so finally we got spectra with resolutions of 23.26 \AA/pixel and 77.35 \AA/pixel respectively (with a
redshift of 2.5; this is equivalent to 6.6 and 22.1 \AA/pixel at rest), 
with average S/N of 3.0 and 2.2 respectively. The spectra were analyzed following the standard procedure using IRAF which comprises bias subtraction, flat-field correction, co-addition of the three single exposures, spectral extraction, wavelength calibration, and flux calibration. The spectroscopic standard star Ross 640 (Oke 1990) was observed for the spectrophotometric calibration in the two nights of the observations.

\begin{figure}
\vspace{1cm}
\centering
\includegraphics[width=8cm]{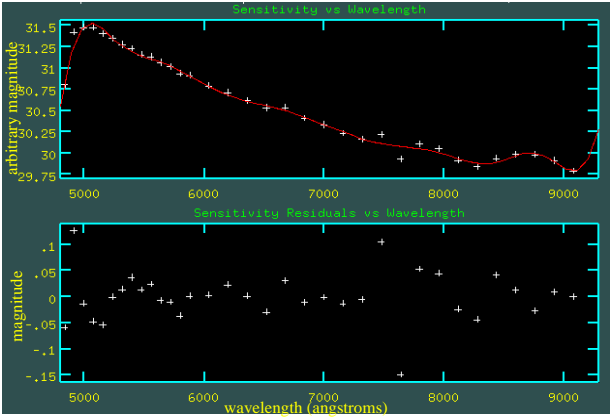}
\caption{Sensitivity versus wavelength in the flux calibration of spectra using the standard star
Ross 640.}
\vspace{1cm}
\label{Fig:fluxcalib}
\end{figure}

The accuracy of the flux calibration is important here because we will fit the shape of the continuum of the spectrum and not its lines, given the low S/N of
the data. We only used the spectral region
between 4980 and 8850 \AA , where the transmission is high enough and allows an error of $<5$\% in the flux absolute calibration, as estimated from possible variations in the way to extract the spectrum (with the command 'apall' of IRAF) and the calibration with the standard star (with the commands 'standard', 'sensfunc', and 'calibrate' of IRAF). We consider this error in the flux calibration negligible for the determination of the error of the parameters to be fitted in Section \ref{.fitting}. In Fig. \ref{Fig:fluxcalib},
we show the accuracy in the response calibration using the standard star Ross 640.
The absolute flux calibration of the spectra was renormalized in order to fit their AB magnitudes in R: 23.60 and 24.89 respectively, and later corrected for Galactic extinction
$A_\lambda \propto \lambda ^{-1.4}$; $A_R=0.047$ and $A_R=0.045$ (Schlafly \& Finkbeiner 2011) for the galaxies \#40812 and \#78891 respectively.

\section{Models}
\label{.models}

\subsection{BC03 model}

The stellar population synthesis by BC03 (also called GALAXEV)
computes the spectral evolution of stellar populations for ages between  $10^5$ and $2\times 10^{10}$ yr at a resolution of 3 \AA \ along the wavelength range from 3200 to 9500 Å for a range of metallicities [Z/H] between -0.40 and 0.40. These predictions are based on the observed stellar library STELIB (Le Borgne
et al. 2003), which allows a reasonable coverage of metallicities around solar. It also computes the spectral evolution across a larger wavelength range, from 91 \AA \ to 160 $\mu $m, at lower resolution using theoretical stellar spectra. 
Padova 1994 evolutionary tracks (Bertelli et al. 1994) and Chabrier (2003) IMF (initial mass function) were used.
The model incorporates all the stellar evolutionary phases and reproduces well the observed optical and near-infrared color-magnitude diagrams of Galactic star clusters of various ages and metallicities, and typical galaxy spectra from the early data release (EDR) of the Sloan Digital Sky Survey (SDSS) (see BC03 for more details). 
Here we use the 30 templates (ten ages, three metallicities) 
of instantaneous-burst models used by
Tremonti (2003) to fit the continua in SDSS galaxy spectra.

\subsection{E-MILES}
 
We also use the UV-extended E-MILES models, which cover the spectral range
$\lambda\lambda$ $1680-50000$\,\AA\ at moderately high resolution. The models
employ the Hubble's NGSL (Next Generation Spectral Library) 
space-based stellar library (Gregg et al. 2006), as prepared in
Koleva \& Vazdekis (2012), to compute spectra of single-age, single-metallicity
stellar populations blueward of $3540$\,\AA. Other empirical stellar spectral
libraries are employed for redder wavelengths (R\"ock et al.
2016). The models span the metallicity
range $-1.79\leq[M/H]\leq+0.26$ for ages larger than $30$\,Myr, for a suite of
IMF types with varying slopes. These models provide reasonably good fits to the
integrated colors and most line-strengths of nearby early-type galaxies with different masses (Vazdekis et al. 2016). In the relevant rest-frame spectral
range the resolution of these models is FWHM$=3$\,\AA.

\subsection{TP-AGB effect}

We do not take into account the enhanced contributions from thermally pulsing asymptotic giant branch (TP-AGB) stars, as done by some authors (e.g., Maraston 2005, Bruzual 2007). 
For old stellar components ($\gtrsim 1$ Gyr), the contribution from this evolutionary phase is fading as the red giant branch is developing rapidly; moreover, the rest-frame spectral range of our reddest filter does not extend significantly beyond 1 $\mu $m. For the young populations (a few hundred Myr), the flux is dominated by UV light and it has small contribution from near-infrared wavelengths at rest, but the evolutionary phase of TP-AGB stars has no impact on the UV spectral range. Therefore, we consider this effect as a minor correction that will not alter significantly the results of this paper.

\section{Fitting of parameters with a single stellar population}
\label{.fitting}

Either with E-MILES or BC03, we have a set of normalized spectra that depend on the age and metallicity of the represented quiescent galaxy: $L^{\rm E-MILES/BC03}
({\rm age,[M/H]};\lambda _{\rm rest})$. If the galaxy is at redshift $z$, the corresponding observed flux would be 
\begin{equation}
F^{\rm E-MILES/BC03}({\rm age,[M/H],z};\lambda)=
\end{equation}
\[ \ \ \ \ \ \ \ \ \ \ \ \ \ \ \ 
\frac{L_0L^{\rm E-MILES/BC03}
({\rm age,[M/H]};\lambda /(1+z))}{4\pi d_L(z)^2(1+z)}
,\]
where $d_L(z)$ is the luminosity distance corresponding to a redshift $z$ with a standard cosmology of $\Omega _m=0.3$, $\Omega _\Lambda =0.7$, $H_0=70$ km/s/Mpc. 

With our spectra and photometric data, we fit the free parameters age, [M/H], $z,$ and the amplitude $L_0$ by minimizing the reduced chi-square,
\begin{equation}
\label{chi2red}
\chi^2_{\rm red}=\frac{1}{N_{\rm dof}}\sum _{i=1}^N\frac{[F_{\rm obs.}(\lambda _i)-F^{\rm E-MILES/BC03}({\rm age,[M/H]},z;\lambda _i)]^2}{\sigma ^2(\lambda _i)}
,\end{equation}
where $N_{\rm dof}=N-\nu$ is the number of degrees of freedom for $\nu $ free parameters (in our case here, $\nu =4$); $N$ is the number of available points $\lambda _i$, both in spectroscopy and photometry. The $\sigma _i$ is the error bar of the data:
 it is available for the photometry and calculated for the spectra (we take for all pixels the same $\sigma (\lambda _i)$ of $3.5\times 10^{-19}$ and 
$1.5\times 10^{-19}$ erg/s/cm$^2$/\AA \ for the binned spectra of galaxies \#40812 and \#78891, which correspond to the average S/N of 3.0 and 2.2 measured in these spectra per pixel).
We constrain the range of redshifts to be $z>0.5$, which is reasonable given
our previous measuments of the photometric redshifts.
The values of $F^{\rm E-MILES/BC03}$ which correspond to the observed $\lambda _i$ are obtained by
binning and interpolating the model in the case of spectra, or by convolving the spectra of the theoretical model with the transmission curves corresponding to the broad-band filters for which
the photometry was carried out.

Once we get the free parameters that minimize the previous expression in Eq. (\ref{chi2red}), their error bars are derived by constraining the models that follow $\chi^2_{\rm red}<\chi^2_{\rm red, minimum}
\left[1+\frac{f(CL)}{N_{\rm dof}}\right]$ with $f(CL)=4.7$ for 68.3 C.L. (1$\sigma $) and $f(CL)=9.7$ for 95.4 C.L. (2$\sigma $), taking into account that we use four free parameters (Avni 1976). The fact that we have included a factor $\chi^2_{\rm red, minimum}$ in the right hand of the inequality is to take into account that in
cases with $\chi^2_{\rm red, minimum}\ne 1$ the errors were underestimated or overestimated; that is, this is equivalent to assuming that our fit is a ``good fit'' 
(with $\chi^2_{\rm red, minimum}\approx 1$) and multiplying the error bars by some
factor to get it.
BC03 has the advantage of using the whole spectrum range 
and photometric data, whereas E-MILES models are
constrained in a more limited wavelength range in the UV 
with $\lambda _{\rm rest}>$1\,680 \AA.

\subsection{Results}

\begin{figure*}
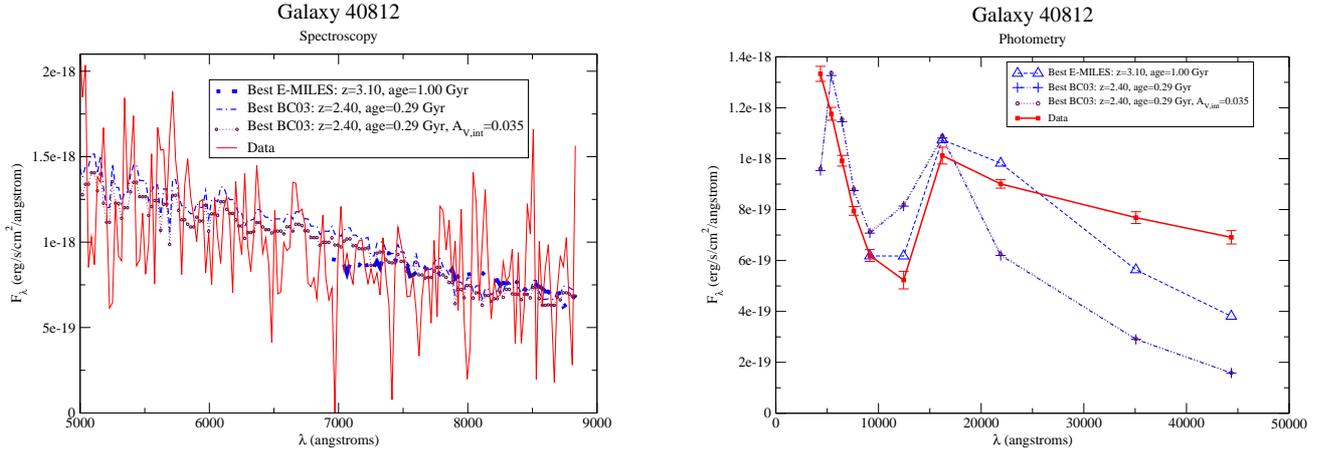

\vspace{1cm}
\centering
\includegraphics[width=8cm]{fitspec.eps}
\hspace{1cm} %
\includegraphics[width=8cm]{fitphot.eps}\\
\caption{Analysis of galaxy \#40812: Best fits to the combination of spectrum and photometric data
obtained with the E-MILES, BC03, and BC03 with internal extinction. 
Left: Spectrum. Right: Photometry.}
\vspace{1cm}
\label{Fig:40812f}
\end{figure*}

\begin{figure*}
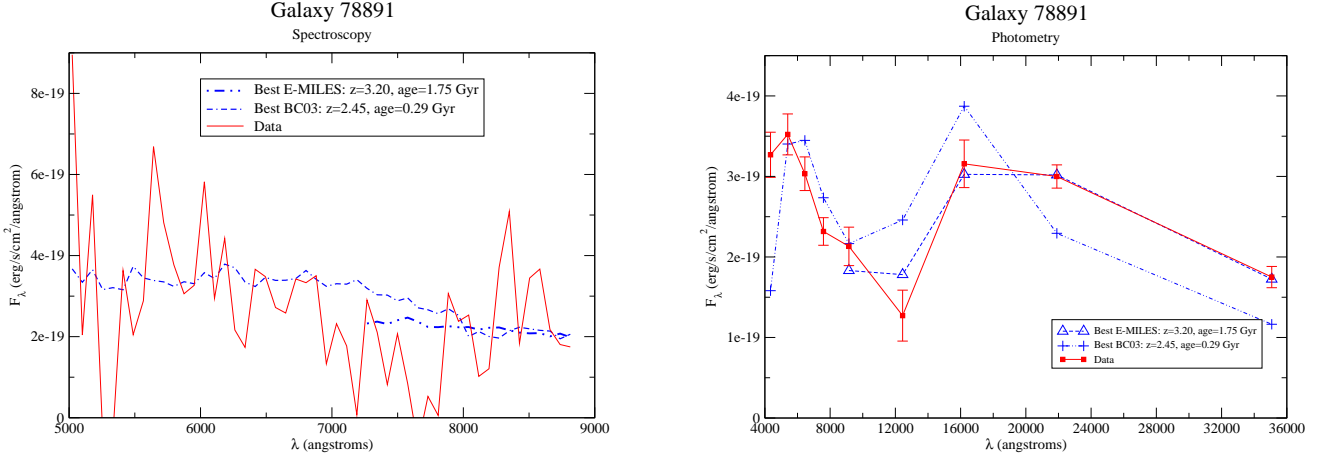

\vspace{1cm}
\centering
\includegraphics[width=8cm]{fitspec2.eps}
\hspace{1cm} %
\includegraphics[width=8cm]{fitphot2.eps}\\
\caption{Analysis of galaxy \#78891: 
Best fits to the combination of spectrum and photometric data
obtained with the E-MILES and BC03.
Left: Spectrum. Right: Photometry.}
\vspace{1cm}
\label{Fig:78891f}
\end{figure*}

The results for the two galaxies for which we obtained the spectra are given in Table \ref{Tab:1SSP} and in Figs. \ref{Fig:40812f} and \ref{Fig:78891f}. The metallicities are almost without constraint from the fits, which
indicates that the shape of the spectra is more sensitive to the age than to the metallicity, as expected in this age regime. There are no big breaks around 2\,000 \AA, which indicates that the galaxies cannot be too old (a galaxy with 2 Gyr would produce a conspicuous break, which is not observed here); although there is some dispersion among the results for the redshift and age of the galaxies with the different fits, they are roughly compatible with each other.

 The ages correspond to an intermediate age population. 
By averaging the results of both models (the data are not independent so the error bars are not reduced proportionally to $\sqrt{N}$ by combining $N$ data), we obtain 
$0.90^{+0.10}_{-0.12}$ Gyr and $0.37^{+0.48}_{-0.30}$ Gyr respectively for galaxies \#40812 and \#78891. 

With the same kind of combination, the most likely redshifts are
$2.54\pm 0.12$ for  galaxy \#40812 and $2.48\pm 0.21$ 
for galaxy \#78891.
Other photometric redshifts have been published: for galaxy \#40812:
$z=2.57$ (CL12), $z=2.63$ (Rowan Robinson et al. 2008), $z=2.16$ (Rowan Robinson et al. 2013), $z=2.53$ (Williams et al. 2009); for galaxy \#78891:
$z=2.71$ (CL12), $z=2.38$ (Rowan Robinson et al. 2008), $z=2.23$ (Rowan Robinson et al. 2013), $z=2.94$ (Williams et al. 2009). They are roughly compatible with our results here, provided that we allow a generous error bar for these numbers (up to 20\%).

Nonetheless, the fits with single stellar population for the range of photometric and spectroscopic data are not good. Only the fit with E-MILES to galaxy \#78891 is
good because it finds a solution at high $z$ compatible with the highest wavelength data, but this solution fails again when we extend the data to the far UV region using the BC03 models. We see that it is not possible to reconcile both the infrared photometry with the ultraviolet data: a good fit to the shortest wavelengths does necessarily
lead to a flux prediction which is too low in the mid-infrared. This denotes that the galaxies are
either more complex than quiescent galaxies with a single episode of stellar formation, or that there is some non-negligible extinction. We will explore these two scenarios in the following sections.

   \begin{table*}
      \caption[]{Best fit parameters obtained with a single stellar population for galaxies \#40812 and \#78891 combining spectroscopy and photometry.}
         \label{Tab:1SSP}
         \begin{tabular}{lccccc}
            \hline
            \noalign{\smallskip}
Galaxy & Model & $\chi ^2_{\rm red.}$
& $z$ & Age (Gyr) & [M/H]  \\
            \noalign{\smallskip}
            \hline
            \noalign{\smallskip}
\#40812 & E-MILES & 3.69 & $3.10^{+0.20}_{-0.20}$ &
$1.00^{+0.75}_{-0.10}$ & $-1.26^{+0.30}_{-1.01}$ \\ 
\#40812 & BC03 & 9.51 & $2.40^{+0.10}_{-0.10}$ &
$0.29^{+0.34}_{-0.19}$ & $-0.40^{+0.40}_{-0}$ \\ 
\#78891 & E-MILES & 1.07 & $3.20^{+0.25}_{-0.70}$ &
$1.75^{+1.00}_{-1.25}$ & $-1.79^{+1.54}_{-0.48}$ \\ 
\#78891 & BC03 & 3.33 & $2.45^{+0.15}_{-0.15}$ &
$0.29^{+0.34}_{-0.19}$ & -0.4, unconstr. \\  \hline
            \noalign{\smallskip}
            \hline
         \end{tabular}
   \end{table*}

\subsection{Adding extinction}
\label{.ext1}

In our fits, we have assumed that our galaxies are dust free following the
CL12 selection based both on color-color diagnostic and on 
low 24 $\mu $m emission. 
At 2175 \AA \ at rest, we do not see any bump in the spectra, 
which would be characteristic of a significant extinction in a Milky Way-like object. 
We tested whether the bad fits obtained with the BC03 model using only one SSP could be improved if we allow some internal extinction in our galaxies.
As we have seen that metallicity does not
play an important role in fits, we substitute it with a new free parameter: the internal extinction normalized in V at-rest $A_{V,int}$. In this case we use the value of
metallicity of the previous fit.
For the extinction law, we use
an attenuation curve often used in high-redshift studies: 
the law derived by Calzetti et al. (2000), derived empirically for a sample of nearby starburst (SB) galaxies. The 2175 \AA \ bump is absent for this law: this characteristic suggests that starburst galaxies contain small magellanic cloud-like dust grains rather than Milky Way-like grains. 
The application of this law is suggested for the central star-forming regions of galaxies, and therefore is appropriate for high-$z$ galaxies. For this reason, Calzetti's law is used to correct the value of the SFR at high-redshift, and we use this extinction law here, with the parameter $R_V=4.05$.
Then, we perform our fit with three free parameters, $z$, age, and $A_{V,int}$, for the case of the fit of \#40812 with BC03 model with fixed [M/H]=-0.4.
The best fit is obtained for $z=2.40$, age=0.29 Gyr, and $A_{V,int}=0.035$, getting a $\chi ^2_{red.}=9.47$, not much better than the fit without extinction and free
metallicity (which provided $\chi ^2_{red.}=9.51$); this is illustrated in Fig. \ref{Fig:40812f}. We conclude that the obtained
mismatch cannot be attributed to internal extinction
in the galaxies.

\section{Fitting of parameters with two single stellar populations}
\label{.fit2}

Here we perform the fits using two single stellar populations (SSPs), one of them relatively old and another one young.  
We allow five free parameters, apart from the amplitude: the redshift ($z$), the ratio of old/young population ($A_2$), the age of the young
population $age_1$ (constrained to be less than 0.3 Gyr), the age of the old population $age_2$ , and the metallicity [M/H] (the same for both populations; indeed
for the younger component it is not important what the metallicity is because
the models are almost completely insensitive to it). For $\nu =6$ 
free parameters, $f(CL)=7.0$ for $1\sigma $ or 12.8 for $2\sigma $.

\begin{equation}
F^{\rm BC03/E-MILES}({\rm age_2,A_2,z};\lambda)=
\end{equation}\[ \ \ \ \ \ \ \ \ \ \ \ \ \ \ \
\frac{L_0L^{\rm BC03/E-MILES}
({\rm age _1,[M/H]};\lambda /(1+z))}{4\pi d_L(z)^2(1+z)}
\]\[ \ \ \ \ \ \ \ \ \ \ \ \ \ \ \ 
+\frac{A_2L_0L^{\rm BC03/E-MILES}
({\rm age _2,[M/H]};\lambda /(1+z))}{4\pi d_L(z)^2(1+z)}
.\]

\begin{figure*}
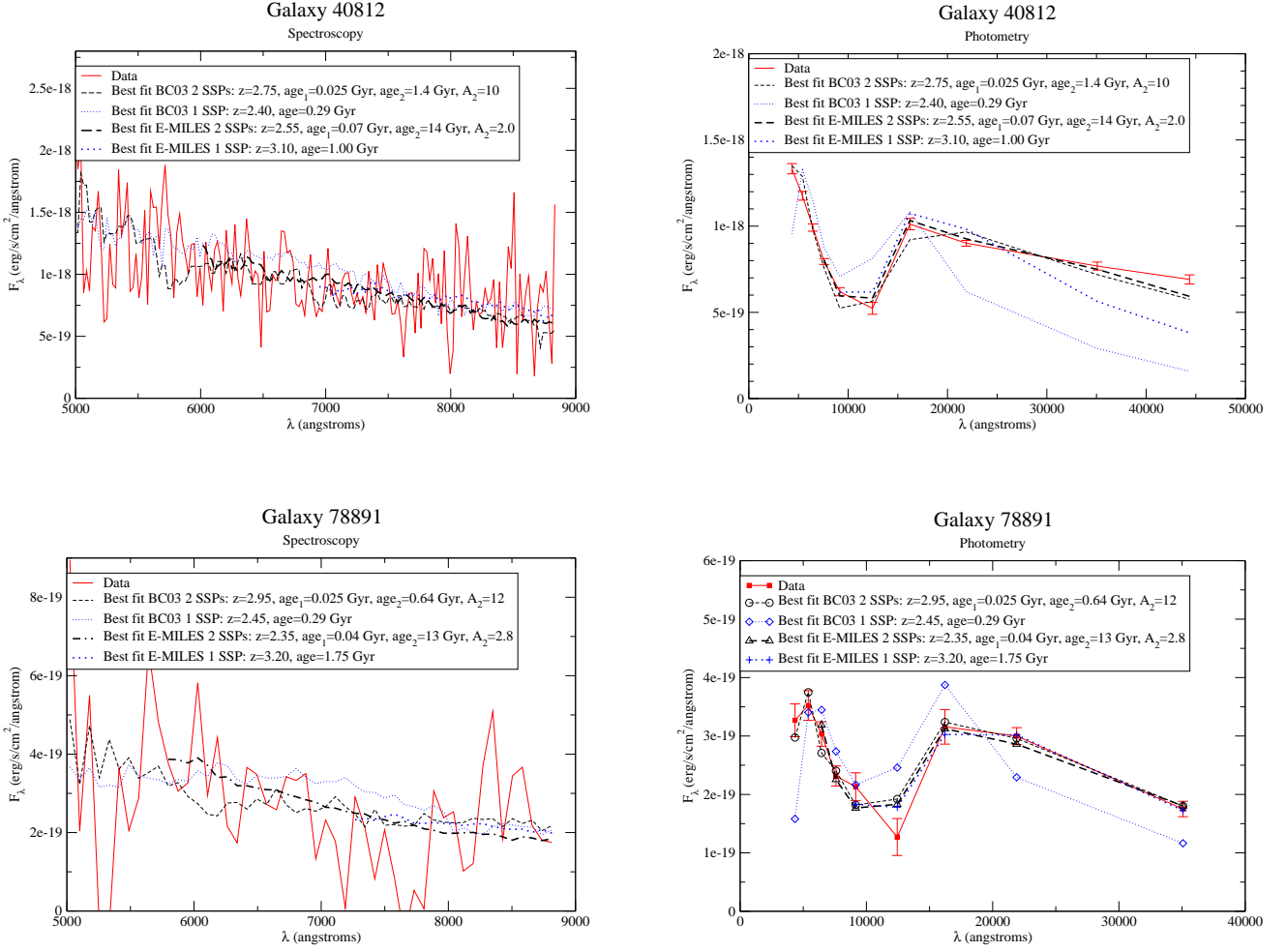

\vspace{1cm}
\centering
\includegraphics[width=8cm]{fitspec_2SSP.eps}
\hspace{1cm}
\includegraphics[width=8cm]{fitphot_2SSP.eps}\\
\vspace{1cm}
\includegraphics[width=8cm]{fitspec2_2SSP.eps}
\hspace{1cm}
\includegraphics[width=8cm]{fitphot2_2SSP.eps}\\
\caption{Spectrum and photometry at observed wavelength corresponding to the best fits with two SSPs (see Table \ref{Tab:2SSP}) or with one SSP (see Table \ref{Tab:1SSP}) of spectroscopy or photometry for galaxies \#40812 and \#78891.}
\vspace{1cm}
\label{Fig:2SSP}
\end{figure*}

\begin{figure*}
\vspace{1cm}
\centering
\includegraphics[width=8cm]{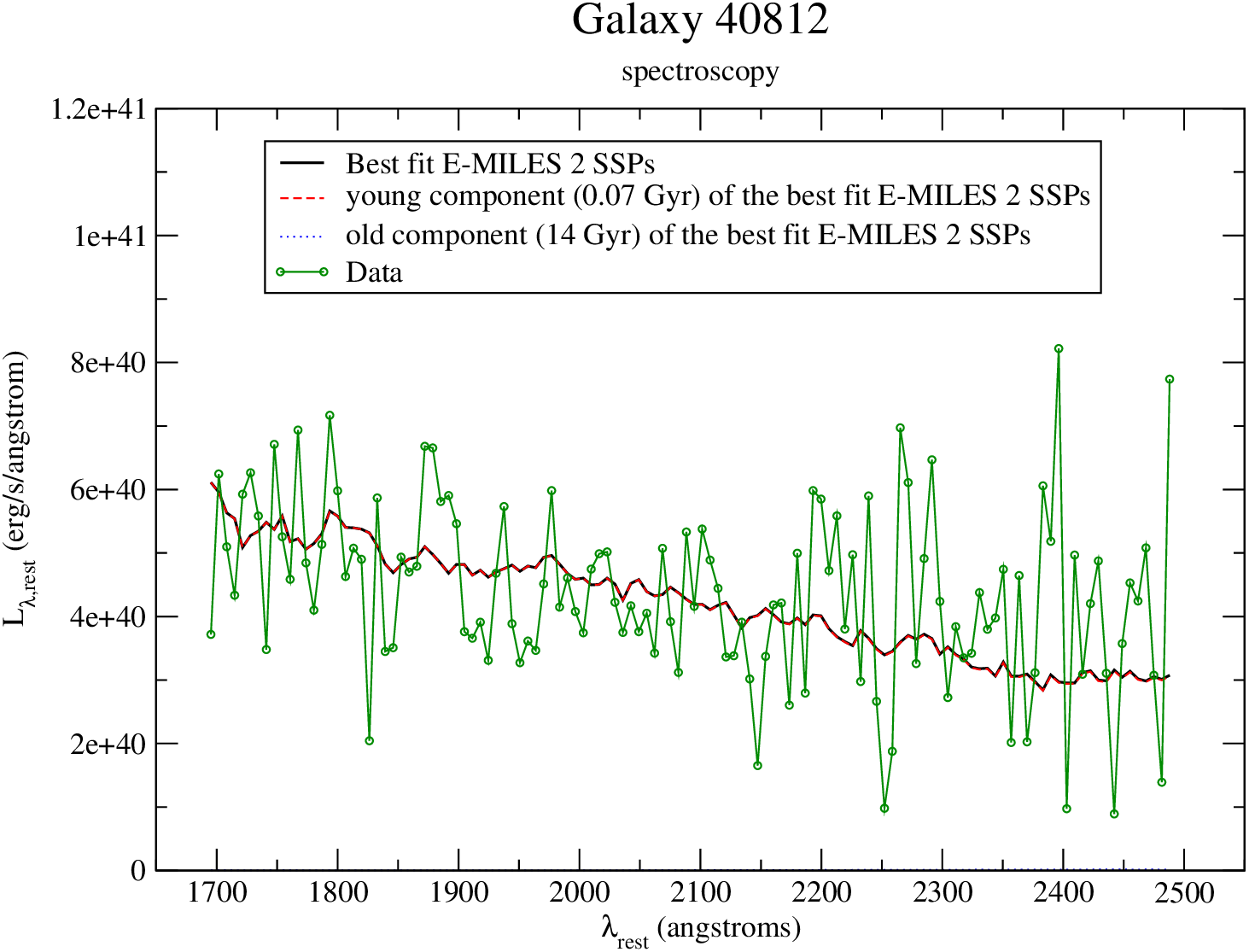}
\hspace{1cm}
\includegraphics[width=8cm]{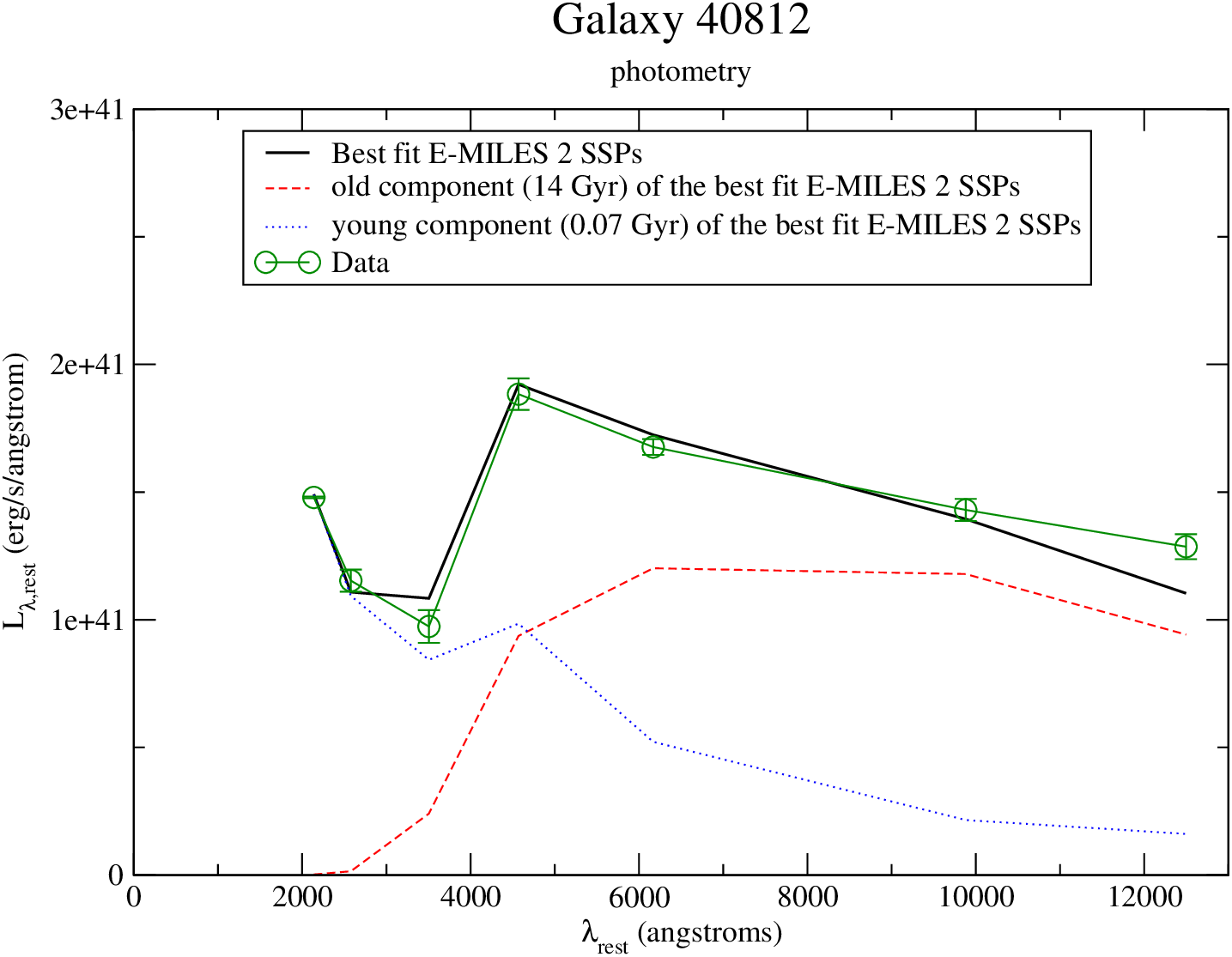}\\
\vspace{1cm}
\includegraphics[width=8cm]{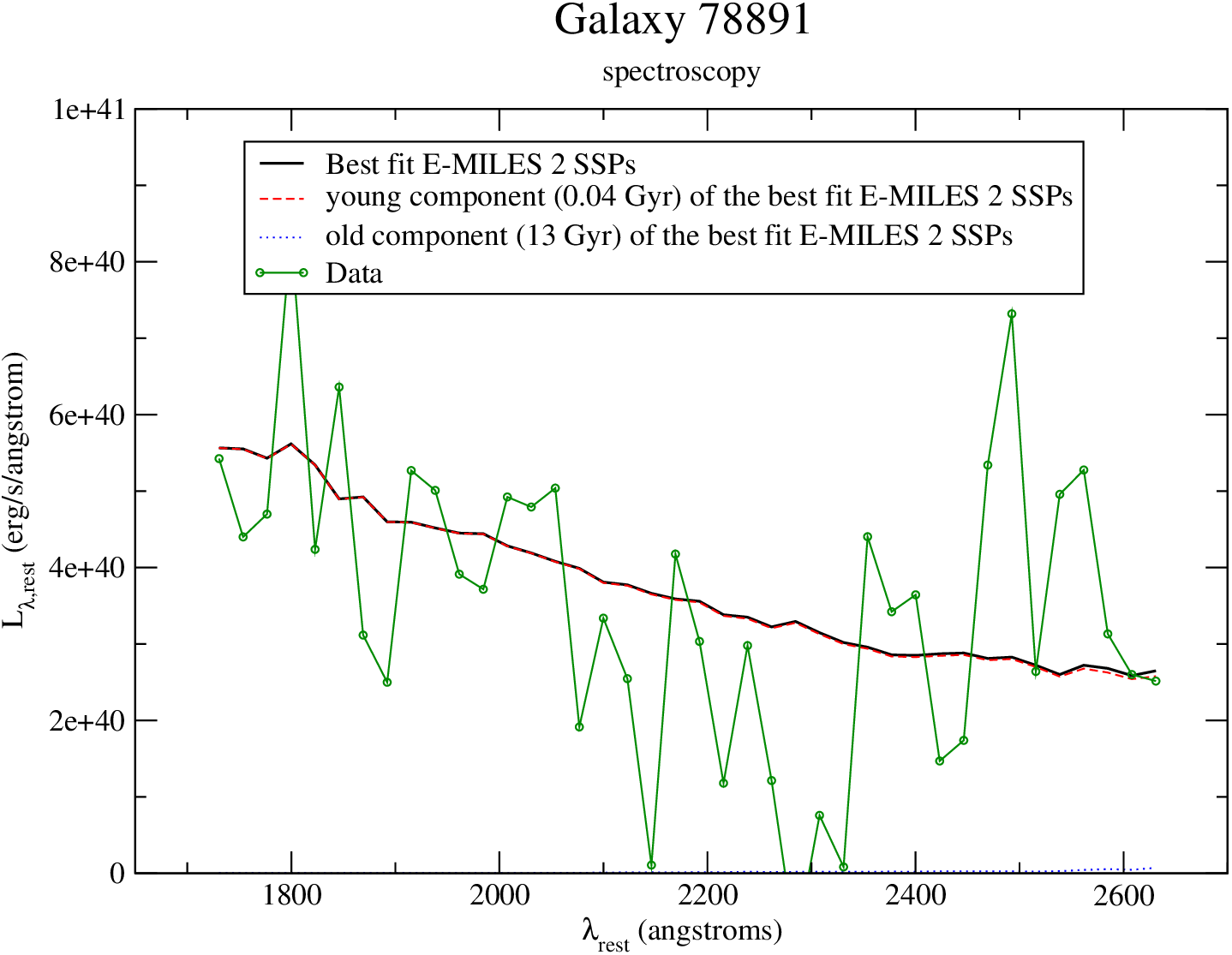}
\hspace{1cm}
\includegraphics[width=8cm]{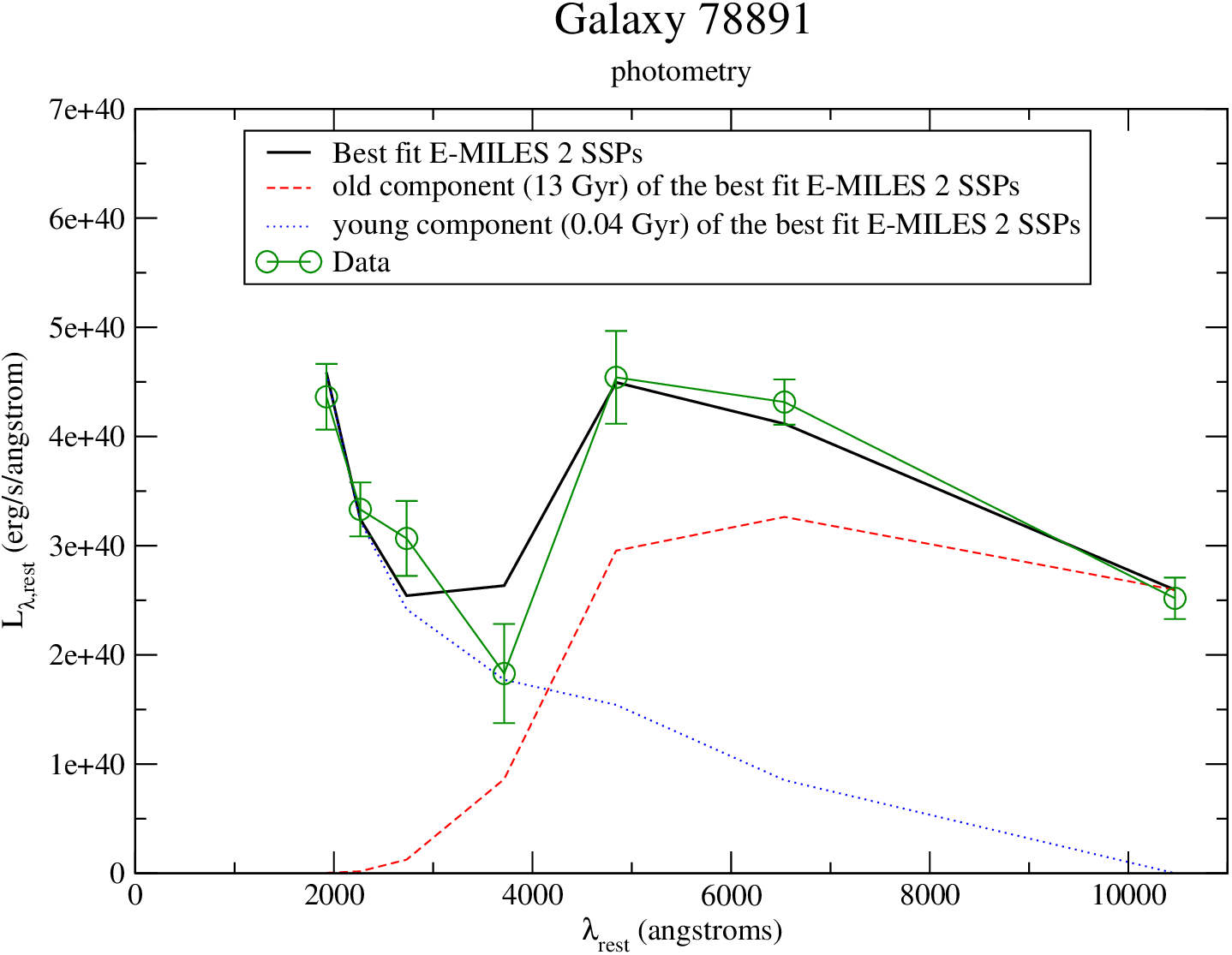}\\
\caption{Spectrum and photometry at rest corresponding to the best fits with two SSPs of E-MILES model (see Table \ref{Tab:2SSP}) of spectroscopy or photometry for galaxies \#40812 and \#78891, indicating the decomposition of the young and the old component.}
\vspace{1cm}
\label{Fig:2SSPl}
\end{figure*}

\begin{figure*}
\vspace{1cm}
\centering
\includegraphics[width=8cm]{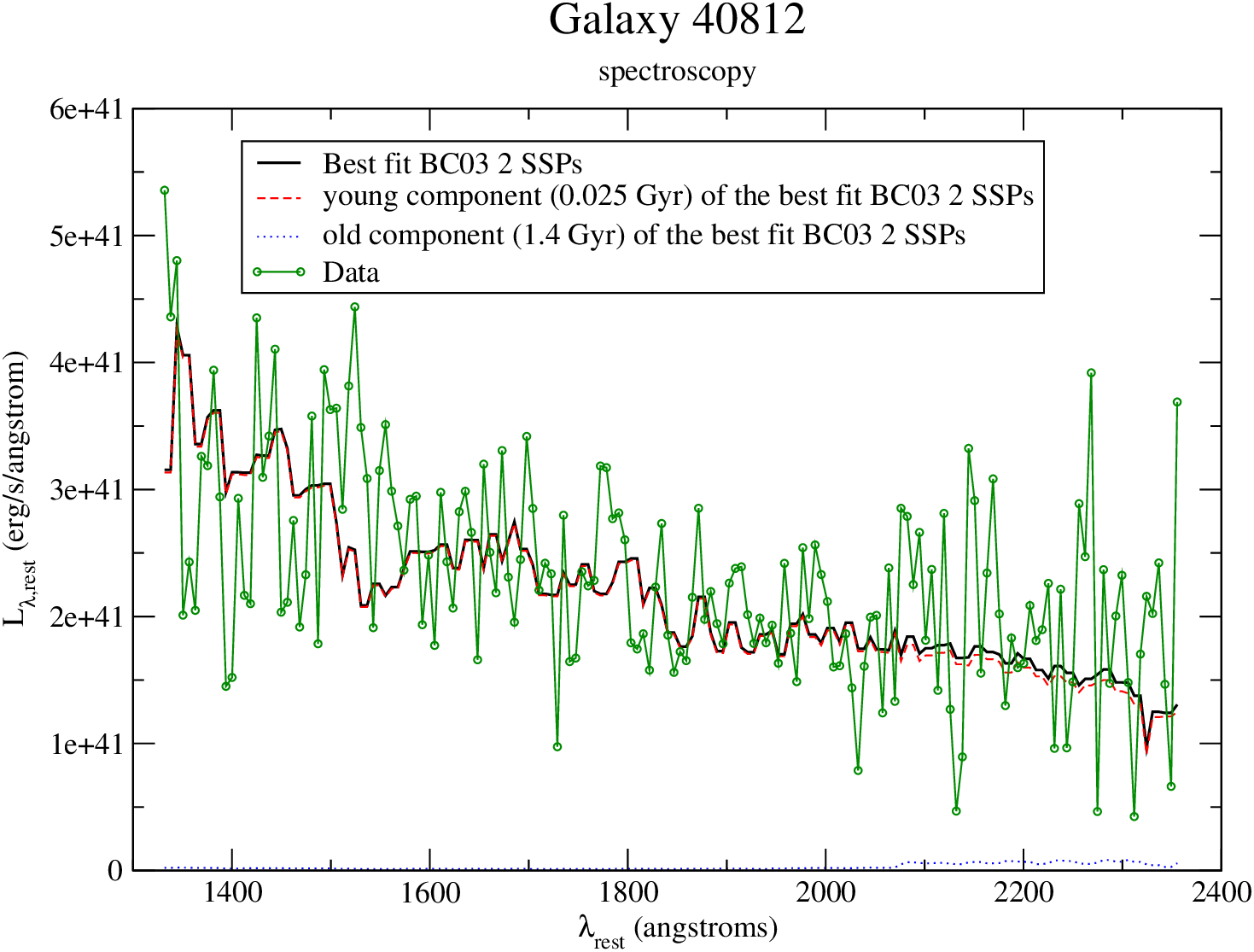}
\hspace{1cm}
\includegraphics[width=8cm]{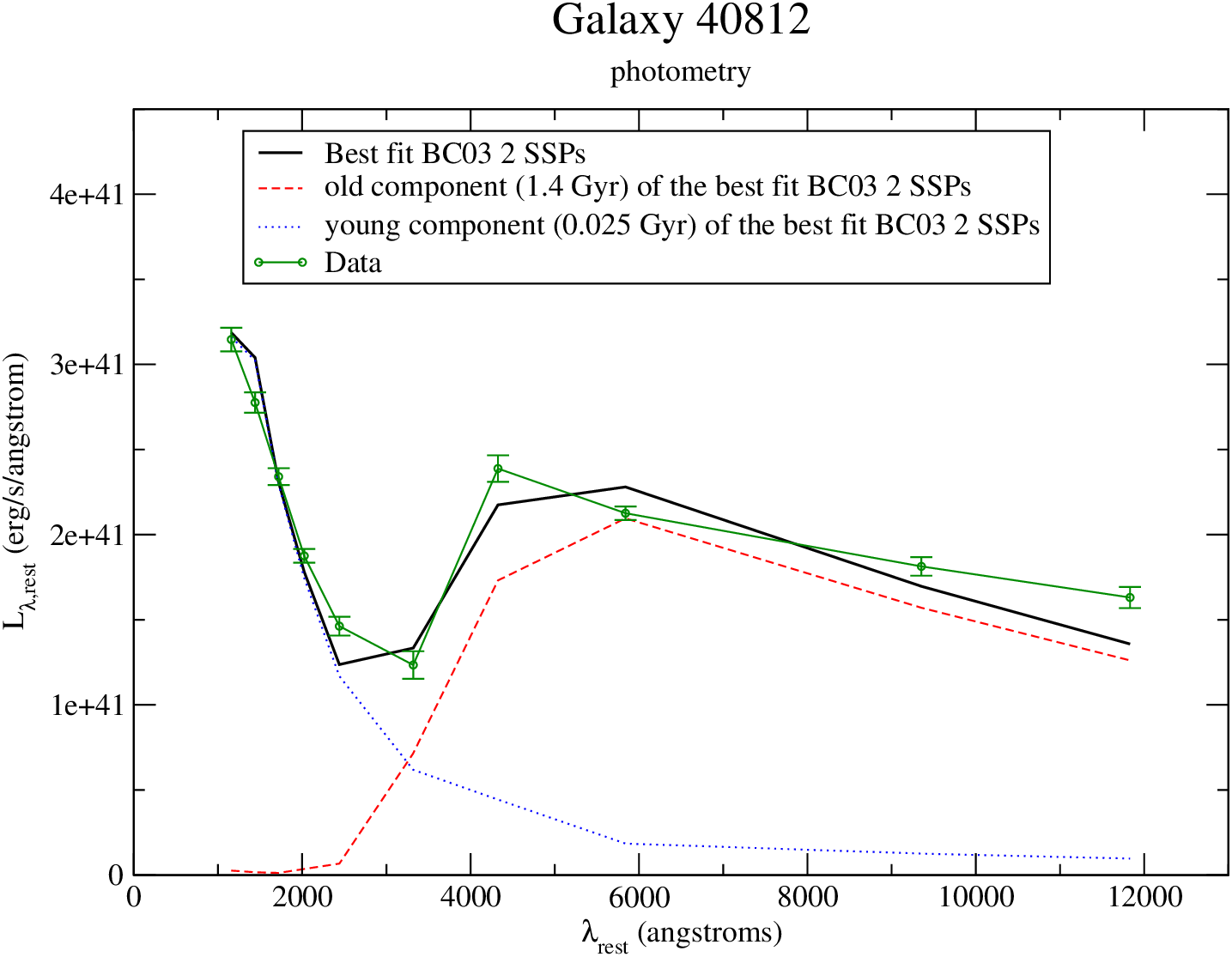}\\
\vspace{1cm}
\includegraphics[width=8cm]{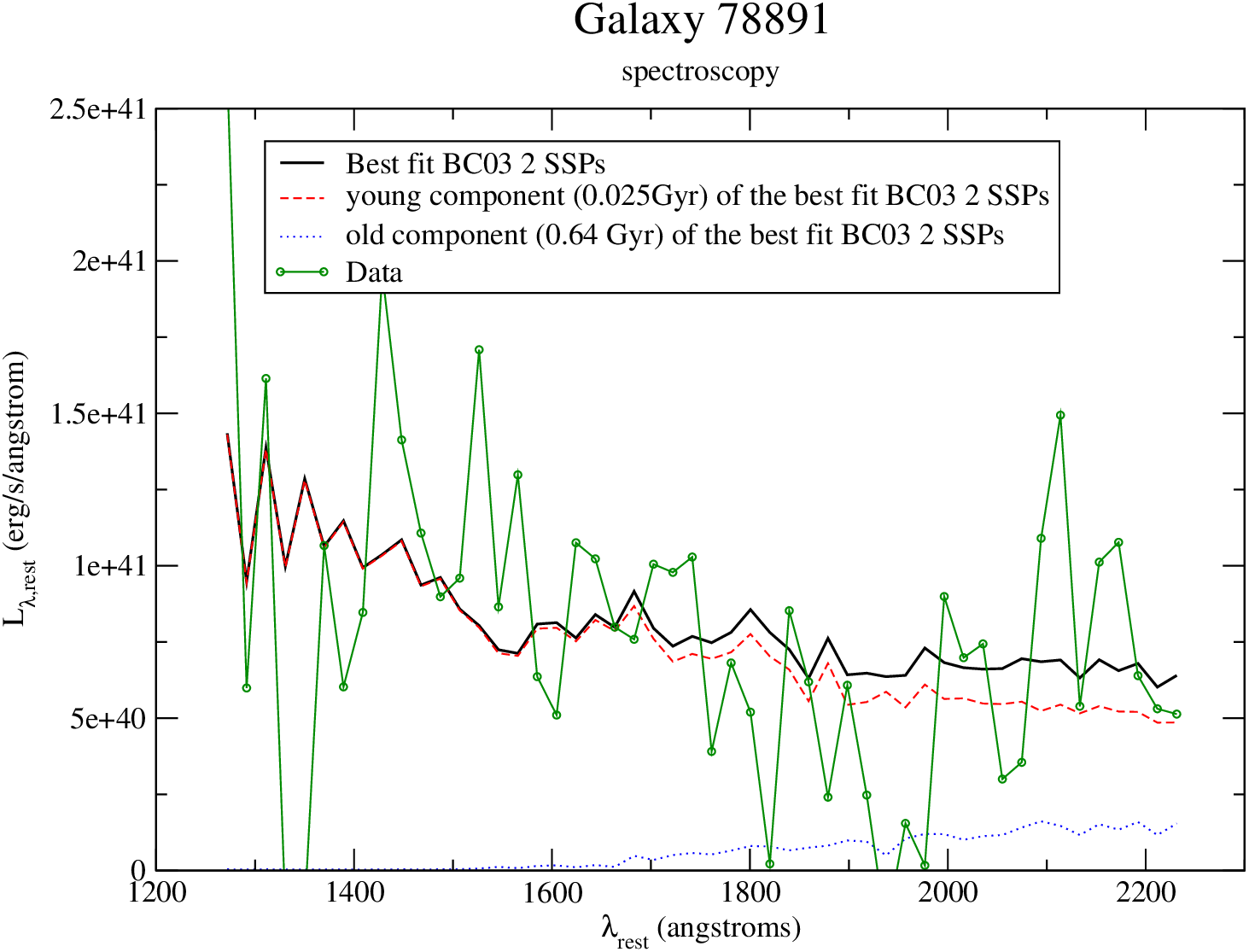}
\hspace{1cm}
\includegraphics[width=8cm]{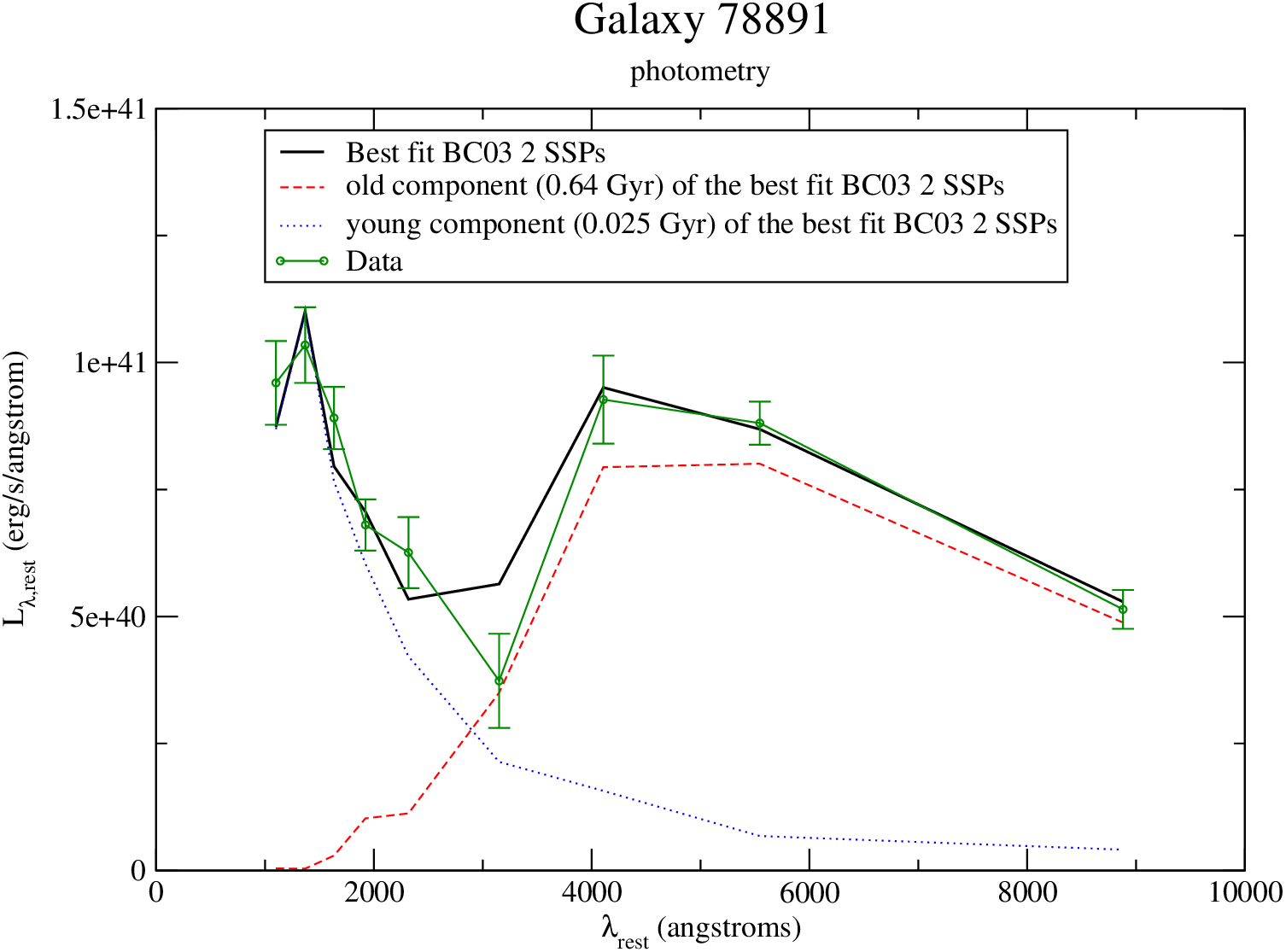}\\
\caption{Spectrum and photometry at rest corresponding to the best fits with two SSPs of BC03 model (see Table \ref{Tab:2SSP}) of spectroscopy or photometry for galaxies \#40812 and \#78891, indicating the decomposition of the young and the old component.}
\vspace{1cm}
\label{Fig:2SSPl2}
\end{figure*}

The normalization of the spectra is such that $L^{\rm BC03/E-MILES}
(\lambda =5500\ \AA)=1$. This means that $A_2$ represents
the ratio of old/young luminosity at 5\,500 \AA \ at rest. 
The result of the fits for both galaxies is given in Table \ref{Tab:2SSP} and in Figures \ref{Fig:2SSP}, \ref{Fig:2SSPl}, and \ref{Fig:2SSPl2}.

   \begin{table*}
      \caption[]{Best fitted parameters with two stellar populations for galaxies \#40812 and \#78891.}
         \label{Tab:2SSP}
         \begin{tabular}{lcccccccc}
            \hline
            \noalign{\smallskip}
Galaxy & Model & $\chi ^2_{\rm red.}$
& $z$ & Age$_1$ (Gyr) & Age$_2$ (Gyr) & $A_2$ & [M/H] & Young/old stellar mass\\
            \noalign{\smallskip}
            \hline
            \noalign{\smallskip}
\#40812 & E-MILES & 0.82 & $2.55^{+0.25}_{-0.10}$ & $0.07^{+0.02}_{-0.01}$ &
$14^{+0}_{-3.5}$ & $2.0^{+0.2}_{-0.4}$ & $+0.40^{+0}_{-0.14}$  & $0.042^{+0.013}_{-0.010}$ \\ 
\#40812 & BC03 & 1.41 & $2.75^{+0.10}_{-0.10}$ & 0.025$^{+0.075}_{-0.020}$ &
$1.4^{+3.6}_{-0.5}$ & $10^{+1}_{-2}$ & $+0.40^{+0}_{-0.40}$  & $0.009^{+0.008}_{-0.009}$ \\ 
\#78891 & E-MILES & 0.83 & $2.35^{+0.60}_{-0.20}$ & 0.04$^{+0.26}_{-0.04}$ &
$13^{+1}_{-12}$ & $2.8^{+7.2}_{-0.4}$ & $+0.15^{+0.25}_{-1.64}$ & $0.005^{+0.014}_{-0.005}$ \\ 
\#78891 & BC03 & 1.34 & $2.95^{+0.05}_{-0.25}$ & 0.025$^{+0.265}_{-0.020}$ &
$0.64^{+0.76}_{-0.35}$ & $12^{+4}_{-3}$ & +0.4, unconstr. & $0.014^{+0.035}_{-0.014}$ \\ \hline
            \noalign{\smallskip}
            \hline
         \end{tabular}
   \end{table*}

The fits with two SSPs are much better (they have lower average reduced $\chi ^2$ ), 
especially when using the BC03 model, which allows us to see that the far UV, 
$\lambda _{\rm rest}<1\,680$ \AA, cannot be
fitted simultaneously with the mid-infrared data for a single stellar population. 
It is clear that at least two
components are necessary.

The plausibility of this second solution with two SSPs may be related to
the possibility that these galaxies are experiencing some starbursts. In principle, our galaxies were
selected to be passively evolving without dust and without starbursts, but we may wonder whether a very small number of young stars is possible. 
The ratio of young to old stellar masses
can be calculated for the luminosity ratios $A_2$ and using 
the mass-luminosity ratio (Vazdekis et al. 2016; assuming 
Kroupa Universal IMF) at a wavelength
of 5\,500 \AA . They are given in Table \ref{Tab:2SSP}. In all cases,
the young stellar mass is less than 5\% of the old stellar mass. This is compatible
with the fact that they are elliptical galaxies dominated in mass by an old population.
The average redshifts with two SSPs (combining MILES and BC03 results) are $2.69\pm 0.15$ and $2.72\pm 0.28$, similar to the redshifts obtained with one SSP.

Again, as stated in Section \ref{.ext1}, we can test whether a modest amount of internal extinction in the galaxies, associated to the young population, may alter significantly the results. The fact that the young population fits the ultraviolet part very well without internal extinction correction indicates that we may most likely neglect extinction at all wavelengths. A small amount of extinction might affect slightly the  young population but has a negligible impact on the old
population that dominates the reddest wavelengths.

To test these effects, we carried out
again the fit of galaxy \#40812 using the BC03 model (which reaches shorter wavelengths in the UV, thus allowing us to be more restrictive in the extinction magnitude), for three values of internal extinction, $A_{V,int}=0.02,0.04,0.08,$ using again Calzetti et al. (2000)'s extinction law.
The result is: $\chi ^2_{red.}=1.43$, 1.48, and 1.65 respectively, slightly worse than
1.41 without extinction, and with similar values of the parameters for the best fits. 
The best fit is obtained without extinction and,
even for a small amount of reddening, 
it does not alter our results.

\section{Fit with extended star formation}
\label{.fitother}

We also tried another approach for fitting the data, which consists
of an exponentially decaying star formation ($\tau $ model,
as used by other authors, e.g., S\'anchez-Dom\'\i nguez et al. 2016).
We may model the global spectral energy distribution as:
\begin{equation}
L(\lambda _{\rm rest})=L_0\int _0^{{\rm age}}dt\ L_*(t,[M/H];\lambda _{\rm rest})e^{-\frac{{\rm age}-t}{\tau }}
,\end{equation}
where $L_*$ is the luminosity per unit mass; 
we only use E-MILES this time, given that it has a higher resolution of ages.
We discretize the integral into a sum of the different available
components.  
A fit of this type, with $z$, $age$, [M/H], and $\tau $ as free parameters gives: for \#40812,
$\chi ^2_{\rm red}=2.62$, $z=3.00^{+0.20}_{-0.10}$,
age=11.5$^{+2.5}_{-4.5}$ Gyr, $\tau=\infty$ (unconstrained within 1$\sigma $) (constant
star formation rate), [M/H]=-0.35 (unconstrained within 1$\sigma $); for \#78891,
$\chi ^2_{\rm red}=1.02$, $z=2.75_{+0.45}^{-0.25}$,
age=14.0$^{+0}_{-11.5}$ Gyr, $\tau=160$ Gyr (unconstrained with 1$\sigma $)
(almost constant star formation rate), [M/H]=-0.66 (unconstrained with 1$\sigma $).
It is remarkable that the best fits obtained here are worse than those achieved with two SSPs ($\chi ^2_{red.}\ge 20$\% larger), although better than with one SSP ($\chi ^2_{red.}\ge 30$\% smaller). We conclude that the model with two SSPs is a better representation of the data, which further supports the residual star formation scenario.

\section{Fitting the other 18 galaxies with 2 SSPs 
using only photometry}
\label{.others}

Table 1\footnote{There is an error in Table 1 of that paper, as columns 5 and 6 were disordered and did not match the information of the galaxies in the first four columns. The corrected Table 1 is at:
http://cdsads.u-strasbg.fr/NOTES/2012A+A...537A..31C.html.} 
of CL12 gives a list of 20 galaxies classified as extremely red objects and old galaxies. We have analyzed two of them in this paper with spectroscopic and photometric data. For the remaining
18 galaxies we only have photometric data. 
We showed in Section \ref{.fit2} that photometry alone shows the need for two SSPs to produce a good fit. Therefore in this section we perform the fits of
these 18 galaxies relying only on their photometric data. Here we only use the BC03 model, since we have seen that both
E-MILES and BC03 give similar results, and BC03 has the advantage over E-MILES of including the wavelength range $\lambda _{\rm rest}<1\,680$ \AA . Applying the same fits as in Section \ref{.fit2} but only with photometry, we get the results given in Table \ref{Tab:2SSPothers} and Fig.
\ref{Fig:2SSPlothers}.

The analysis of these 18 galaxies gives similar results to those obtained
for the two galaxies studied in the previous sections of
this paper: a combination of a young (0.005-0.29 Gyr) and an 
old (0.64-11 Gyr) component gives an acceptable
fit to our data in most cases. These results show that this star formation pattern is common to most of the galaxies studied here. 
The values of $\chi ^2_{\rm red}$ are higher than the analysis with spectra+photometry because the relative photometric errors are much smaller than the spectroscopic ones.
The old population dominates the red-near infrared range at rest, whereas the young population dominates the UV at rest. The obtained redshifts fall
in all cases within the range 2.3-3.9, as expected. The metallicities are unconstrained
because large variations of metallicity do not strongly change the predictions of the
models.

For 11 of the 18 galaxies (those with numbers 1559, 10585, 11488, 14598, 25371, 26888, 28153, 28841, 39991, 79102, and 85772) we observe that the photometric data have the typical V-shaped pattern with minima around $\lambda _{\rm rest}\approx 3\,000$ \AA , and decrease of flux over 6\,000 \AA \ at rest. This pattern, which cannot be fully matched with a single SSP, can only be reproduced with two SSPs, as shown in
Section \ref{.fit2}. This shows that our two galaxies \#40812 and \#78891 analyzed previously are not peculiar cases in their stellar content.

\begin{table*}
      \caption[]{Best fitted parameters with two stellar populations for the 20 galaxies of the CL12 list, first column indicating the number of source in the ultra deep survey XMM-LSS, except \#40812 and \#78891, using only photometry and the BC03 model.}
         \label{Tab:2SSPothers}
         \begin{tabular}{lcccccc}
            \hline
            \noalign{\smallskip}
Galaxy & $\chi ^2_{\rm red.}$
& $z$ & Age$_1$ (Gyr) & Age$_2$ (Gyr) & $A_2$ & [M/H] \\
            \noalign{\smallskip}
            \hline
            \noalign{\smallskip}
\#1559 & 2.13 & $2.3^{+1.4}_{-1.0}$ & $0.10^{+0.19}_{-0.095}$ &
$5^{+6}_{-4.1}$ & $3.5^{+9.0}_{-1.5}$ & +0.4, unconstr. \\ 
\#6165 & 0.83 & $2.4^{+0.4}_{-0.6}$ & $0.29^{+0}_{-0.265}$ &
$1.4^{+9.6}_{-0.8}$ & $7^{+19}_{-4}$ & +0.4, unconstr. \\ 
\#8303 & 0.19 & $2.7^{+0.2}_{-0.2}$ & $0.29^{+0}_{-0.19}$ &
$2.5^{+2.5}_{-0.9}$ & $2.2^{+0.6}_{-0.4}$ & $+0.4^{+0}_{-0.4}$ \\ 
\#10585 & 2.07 & $2.5^{+1.0}_{-0.4}$ & $0.10^{+0.19}_{-0.075}$ &
$11^{+0}_{-9.6}$ & $3.5^{+6.5}_{-1.0}$ & +0.4, unconstr.  \\ 
\#11488 & 2.91 & $2.5^{+1.0}_{-0.4}$ & $0.10^{+0.19}_{-0.075}$ &
$2.5^{+8.5}_{-1.9}$ & $4.5^{+7.5}_{-4.0}$ & +0.4, unconstr.  \\ 
\#14598 & 9.40 & $3.1^{+0.6}_{-0.8}$ & $0.10^{+0.19}_{-0.095}$ &
$0.9^{+10.1}_{-0.875}$ & $2.4^{+7.6}_{-2.4}$ & -0.4, unconstr.  \\ 
\#24756 & 6.71 & $2.5^{+1.0}_{-0.6}$ & $0.29^{+0}_{-0.265}$ &
$11^{+0}_{-10.71}$ & $3.2^{+6.8}_{-2.8}$ & 0, unconstr.  \\ 
\#25371 & 2.18 & $2.3^{+1.2}_{-0.6}$ & $0.10^{+0.19}_{-0.075}$ &
$2.5^{+8.5}_{-1.9}$ & $2.8^{+5.4}_{-2.4}$ & +0.4, unconstr. \\ 
\#26888 & 2.05 & $3.1^{+0.8}_{-1.2}$ & $0.10^{+0.19}_{-0.095}$ &
$11^{+0}_{-10.7}$ & $8.0^{+17.5}_{-5.5}$ & -0.4, unconstr.  \\ 
\#28153 & 3.92 & $2.5^{+0.8}_{-0.4}$ & $0.10^{+0.19}_{-0.075}$ &
$5^{+6}_{-4.1}$ & $4.0^{+5.0}_{-3.0}$ & +0.4, unconstr.  \\ 
\#28841 & 2.85 & $3.3^{+0.8}_{-1.0}$ & $0.10^{+0.19}_{-0.10}$ &
$2.5^{+8.5}_{-2.2}$ & $6.5^{+19}_{-5.5}$ & -0.4, unconstr.  \\ 
\#39991 & 9.06 & $2.9^{+0.6}_{-0.8}$ & $0.10^{+0.19}_{-0.095}$ &
$0.9^{+10.1}_{-0.8}$ & $6.0^{+19.5}_{-6.0}$ & -0.4, unconstr.  \\ 
\#44569 & 0.51 & $2.9^{+0.2}_{-0.4}$ & $0.29^{+0}_{-0.19}$ &
$5^{+6}_{-3.6}$ & $1.0^{+0.8}_{-0.3}$ & +0.4, unconstr. \\ 
\#51547 & 1.90 & $2.7^{+0.8}_{-0.6}$ & $0.29^{+0}_{-0.265}$ &
$2.5^{+8.5}_{-1.9}$ & $3.4^{+6.8}_{-2.8}$ & 0, unconstr.  \\ 
\#65143 & 6.71 & $2.5^{+1.0}_{-0.6}$ & $0.29^{+0}_{-0.265}$ &
$11^{+0}_{-10.7}$ & $3.2^{+7.0}_{-2.8}$ & 0, unconstr.  \\ 
\#79102 & 5.69 & $2.5^{+1.2}_{-1.2}$ & $0.10^{+0.19}_{-0.095}$ &
$11^{+0}_{-10.4}$ & $2.6^{+7.6}_{-1.8}$ & +0.4, unconstr.   \\
\#85772 & 0.93 & $3.1^{+0.6}_{-0.8}$ & $0.10^{+0.19}_{-0.095}$ &
$5^{+6}_{-4.4}$ & $8.5^{+17.0}_{-5.0}$ & -0.4, unconstr.   \\
\#93574 & 3.63 & $3.9^{+1.0}_{-2.6}$ & $0.005^{+0.020}_{-0}$ &
$0.64^{+10.36}_{-0.54}$ & $68^{+\infty}_{-68}$ & -0.4, unconstr.   \\
 \\ \hline
            \noalign{\smallskip}
            \hline
         \end{tabular}
   \end{table*}

\section{Discussion on the age determination}
\label{.discus}

\subsection{The oldest age problem and 
systematic errors in the age that have not been taken into account}
\label{.syst}

Surprisingly, our fits in Section \ref{.fit2} provide some solutions involving ages for the old component older than the Universe: the best fits with E-MILES for galaxies
\#40812 and \#78891 give 14 Gyr and 13 Gyr respectively, although in the second case it is compatible with the age of the Universe within the error bars.
Nonetheless, we must bear in mind that our assumptions (only two stellar populations, without dust, neglecting the light contamination from other galaxies, neglecting the TP-AGB uncertainty at $\lambda _{\rm rest}>7\,000$ \AA, or the assumption of a standard IMF) are only approximations, so there may be additional systematic errors attributed.
Furthermore, in light of the recent evidence for a bottom-heavy IMF in massive early-type galaxies (La Barbera et al. 2013), the oldest component of our fits might be approached by such dwarf-enriched-IMF that redden the spectra by $\sim 0.1$ mag., which is equivalent to the change of slope between 3 and 10 Gyr (Vazdekis et al. 2012).
Therefore, we cannot use the present result to establish firmly
that we have found a galaxy older than the Universe, although certainly it is a very old galaxy.
In fact, the analysis with the model BC03, which comprises the far UV part, gives an
age of 1.4$^{+3.6}_{-0.5}$ Gyr for galaxy \#40812, which is compatible with the age of the Universe. 
Other independent estimators of age for the oldest population are necessary. We note that the oldest population dominates in the reddest wavelengths, so the estimator would be better if it is based on mid infrared observations ($\lesssim 5$ $\mu $m, since the dust emission, although low, may become important over this wavelength; da Cunha et al. 2008, Fig. 5). 

For the age of the oldest population in the remaining 18 galaxies analyzed in
Section \ref{.others}, we cannot set a good constraint. There are nine galaxies whose older stellar population components have best fit ages larger than
3 Gyr, but also lower values are allowed within errors. For the remaining nine galaxies, the best fits for the oldest component are between 0.5-3 Gyr and much lower ages are also allowed within the error bars.

The fact that half of the 20 galaxies of CL12 are better fitted with very high ages 
($\ge 3$ Gyr) is due to the low slope in the spectra of the red-near infrared range at rest, which is almost flat in some cases. Younger populations of
the oldest component require a larger value of $-\frac{dL_\lambda}{d\lambda }$
at $\lambda _{\rm rest}>6\,000$ \AA .
Although we do not know why the reddest filters are providing such
high fluxes, it is clear that the decrease in flux is milder in comparison
to that expected for stellar populations with ages 1-2 Gyr.

\subsection{On the unsuitability of cosmic chronometer methods}
\label{.cosmicchrono}

\begin{figure}
\vspace{1cm}
\centering
\includegraphics[width=8cm]{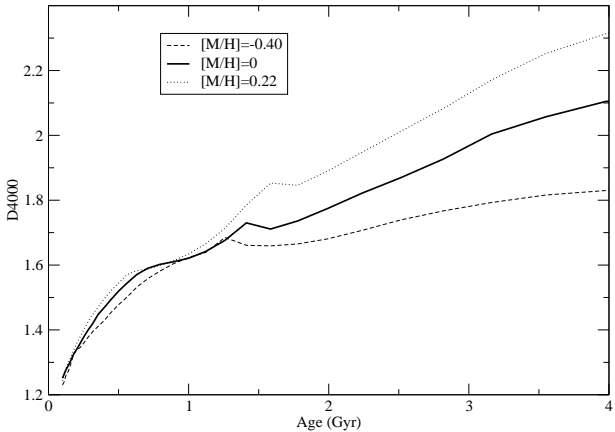}
\caption{Dependence of the Balmer break on the age and metallicity according to the 
E-MILES synthesis model.}
\vspace{1cm}
\label{Fig:D4000}
\end{figure}

Some age-sensitive breaks like the Balmer one at 4\,000 \AA \ at rest 
are contributed by similar light fractions from the two components
(see Fig. \ref{Fig:2SSPl}, left) so they cannot serve as a direct indicator of the age of the oldest population, but instead can indicate an average age of the young and the old population. For instance, if the old population were 4.0 Gyr old and the young population were 0.1 Gyr old, 
with [M/H]=0 both, the Balmer break $D_{4000}\equiv \frac{\overline {F} _\lambda 
(4050-4100 \AA)}{\overline {F}_\lambda (3900-3950 \AA)}$ would be 2.1 and 1.2 respectively
for both populations (using E-MILES model; see Fig. \ref{Fig:D4000}); 
when we mix both populations with a ratio
such that $F_{\lambda ,{\rm old}} (4\,000 \AA)\approx F_{\lambda ,{\rm young}} (4\,000 \AA)$, we get
an average $D_{4000}\approx 1.6$, corresponding to an average age of 0.8 Gyr (using E-MILES model; see Fig. \ref{Fig:D4000}), almost independent of the metallicity. This means that by 
using the Balmer break in these cases and assuming a single population, we underestimate the age of the galaxy.
Such uncertainties pose serious concerns on the reliability of the cosmic chronometer method, used to measure the Hubble parameter as a function of redshift, $H(z)$ (e.g., Moresco et al. 2012, 2016). 

\section{Conclusions}
\label{.conclus}

We present the results of fitting photometric data and spectra 
of two high redshift galaxies employing two different models of stellar populations. The fits of these two types of data with each of these models lead to age, metallicity, and redshift. 
These fits also show that the obtained spectra helped us to reach more constrained results, although it is possible to obtain similar estimates with higher errors based only on the photometric data. The number of points including
spectra is much higher than the number of free parameters and the goodness of the best fit is more accurate. Furthermore, these spectra allowed us to conclude that a young population dominating
the UV wavelengths is required since there is a lack of an abrupt break at 2\,000 \AA \ at rest. Also, the spectra do not show any bump at 2\,175 \AA \ 
characteristic of dust contribution, such as that of the Milky Way. However, we cannot discard the presence of dust contributions like small magellanic clouds or nearby starbursts
(Calzetti et al. 2000).

Models of one SSP do not provide good fits to the photometric data.
We find that combinations of several SSPs are required to simultaneously match both the blue and the red photometric data.
Specifically, we require a dominant component with a very old population and
a small stellar component of $\sim 100$ Myr. 
Remarkably, the mass fraction contribution of the
young component ($<5\%$) 
is in good agreement with the results obtained
by other authors when including the UV spectral range in nearby galaxies
(e.g., Yi et al. 2011, Vazdekis et al. 2016). This is consistent with an
in situ formation scenario with stellar population evolutions evolving
passively. 
We have also tried to fit the data with an
exponentially decaying star formation, 
giving significantly worse results than those achieved with two SSP fits. We note, however, that newer generations of
cosmological hydro-dynamical simulations like EAGLE 
 tend to predict significantly larger star formation histories (Trayford et al. 2015).
The young component of this double SSP is consistent with a residual star formation that can also be observed at lower redshift
galaxies (e.g., Dom\'\i nguez-S\'anchez et al. 2016) and
in the nearby Universe (e.g., Vazdekis et al. 1997, Atlee et al. 2009). This further shows
the unprecedented sensitivity of the UV to the presence of such young
contributions, even for objects at $z>2$. The fits to most of the other
galaxies in our sample, which were performed on the basis of their
photometric data alone, also provide very similar results. 
Our fits favor a scenario where the vast majority of
the stars of these extremely red objects formed rapidly,
with these stellar populations ageing with time, although with small
star formation events from accumulated residual gas.

CL12 gave an average age of 1-2 Gyr for the average population of these kinds of galaxies  based on the relationship of $(B-V)_{\rm rest}$ assuming one SSP, and we see here that there are stars much older than that because we
need at least two SSPs. 
We find that half of the galaxies of our sample have preferred best fits
with ages of the oldest population larger than 3 Gyr, older than the Universe for the standard cosmological model at those obtained redshifts, but statistical and
systematic errors due to the adopted assumptions erase this tension. Further research for these galaxies is necessary.

What is not uncertain and which constitutes the main solid result of this paper 
is the fact that one SSP cannot describe early type galaxies
like the ones analyzed here. It represents a drawback in age estimator methods based on single lines or breaks, such as the Balmer break used in cosmic chronometers, thus invalidating the application
of this technique to constrain the cosmological parameters.

\begin{acknowledgements}
Thanks are given to Helena S\'anchez-Dom\'\i nguez for helpful comments 
on a draft of this paper. We thank the anonymous referee for comments that
helped very much to improve the paper. We would also like to thank the A\&A language editor 
Ruth Chester for the excelent work of revision of this work. 
This work has been supported by the grants  
AYA2013-48226-C3-1-P, AYA2013-48226-C3-2-P, AYA2015-66506-P from the Spanish Ministry of Economy and Competitiveness (MINECO). Based on observations made with the Gran Telescopio Canarias (GTC) installed in the Spanish Observatorio del Roque de los Muchachos of the Instituto de Astrofísica de Canarias on the island of La Palma. 
\end{acknowledgements}

\appendix

\section{Figures of the fit with two SSPs for the 18 galaxies of CL12, only with
photometry}

\begin{figure*}
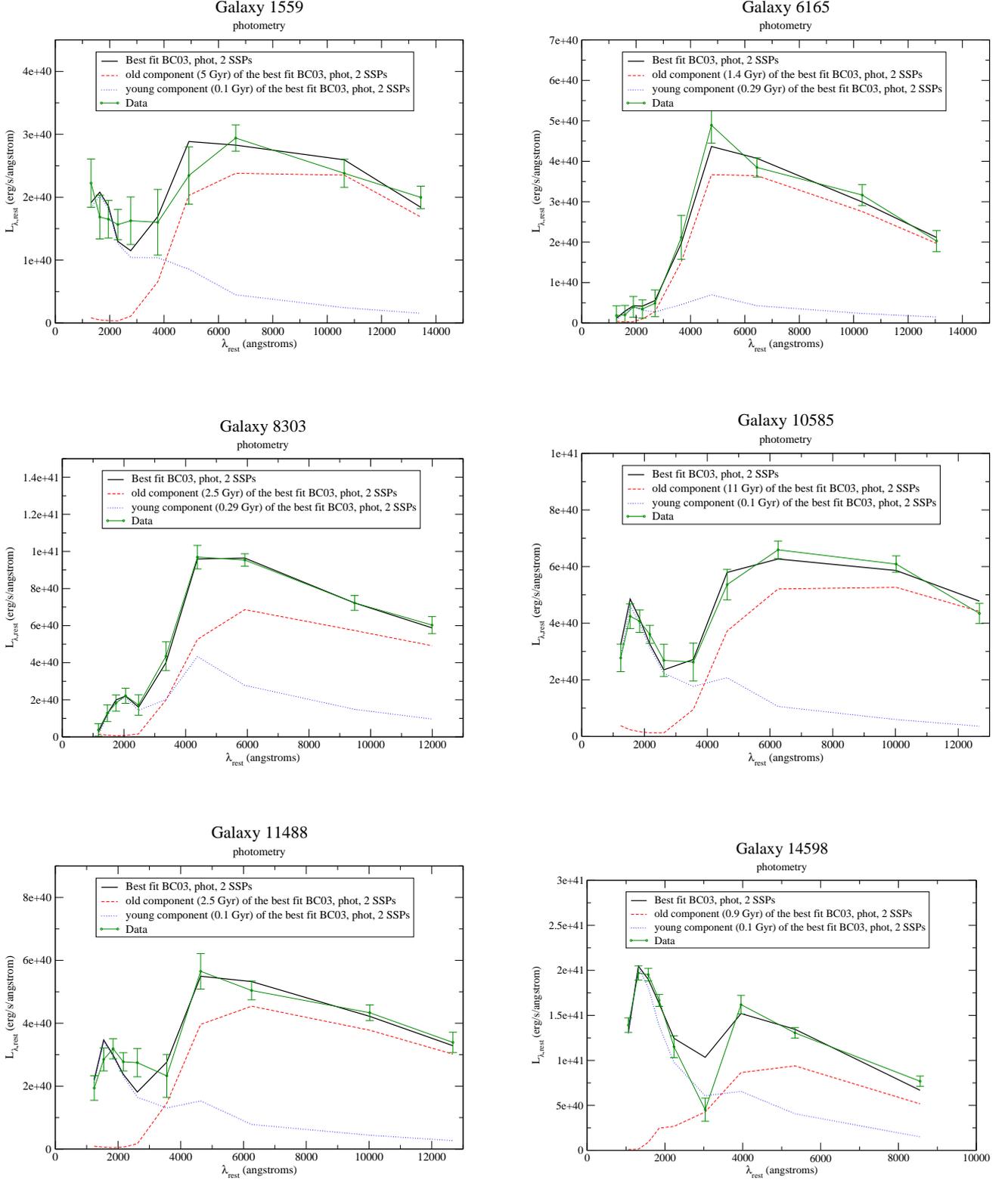

\vspace{1cm}
\centering
\includegraphics[width=8cm]{fitphot3_2SSP_l2.eps}
\hspace{1cm}
\includegraphics[width=8cm]{fitphot4_2SSP_l2.eps}\\
\vspace{1cm}
\includegraphics[width=8cm]{fitphot5_2SSP_l2.eps}
\hspace{1cm}
\includegraphics[width=8cm]{fitphot6_2SSP_l2.eps}\\
\vspace{1cm}
\includegraphics[width=8cm]{fitphot7_2SSP_l2.eps}
\hspace{1cm}
\includegraphics[width=8cm]{fitphot8_2SSP_l2.eps}\\
\caption{Photometry at rest corresponding to the best fits with two SSPs of BC03 model (see Table \ref{Tab:2SSPothers}) for the CL12 galaxies, except \#40812 and \#78891.}
\vspace{1cm}
\label{Fig:2SSPlothers}
\end{figure*}

\begin{figure*}
\vspace{1cm}
\centering
\includegraphics[width=8cm]{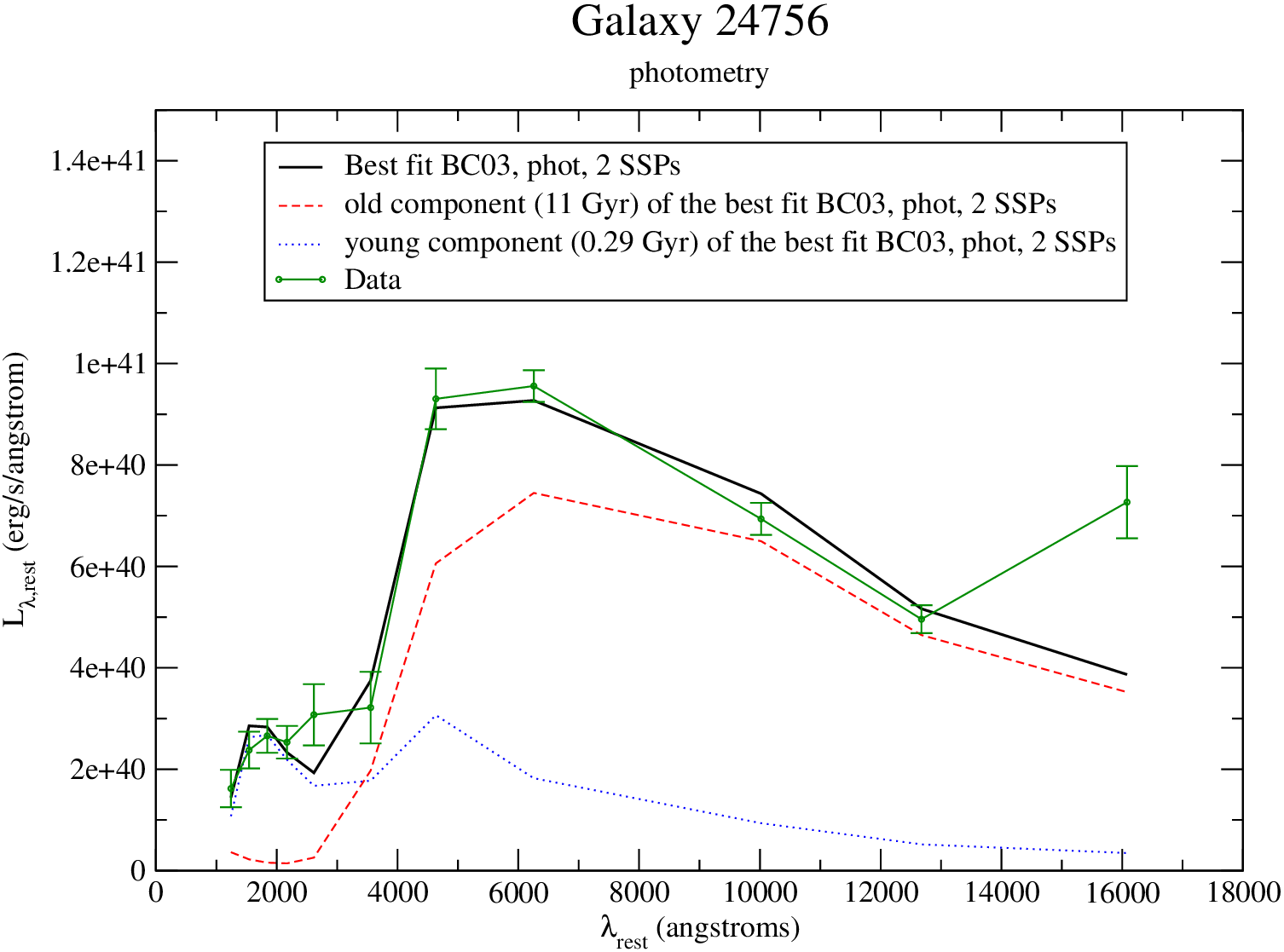}
\hspace{1cm}
\includegraphics[width=8cm]{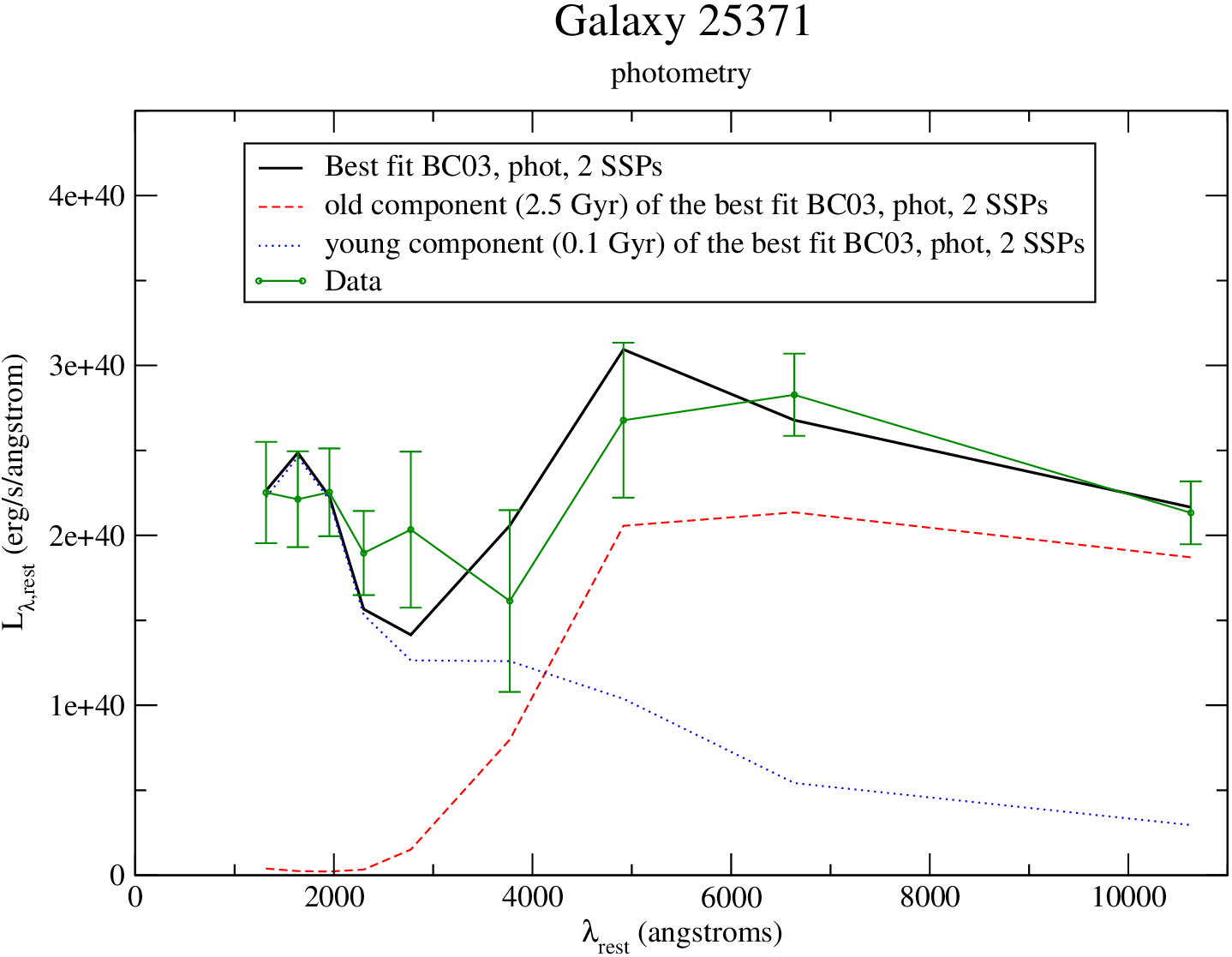}\\
\vspace{1cm}
\includegraphics[width=8cm]{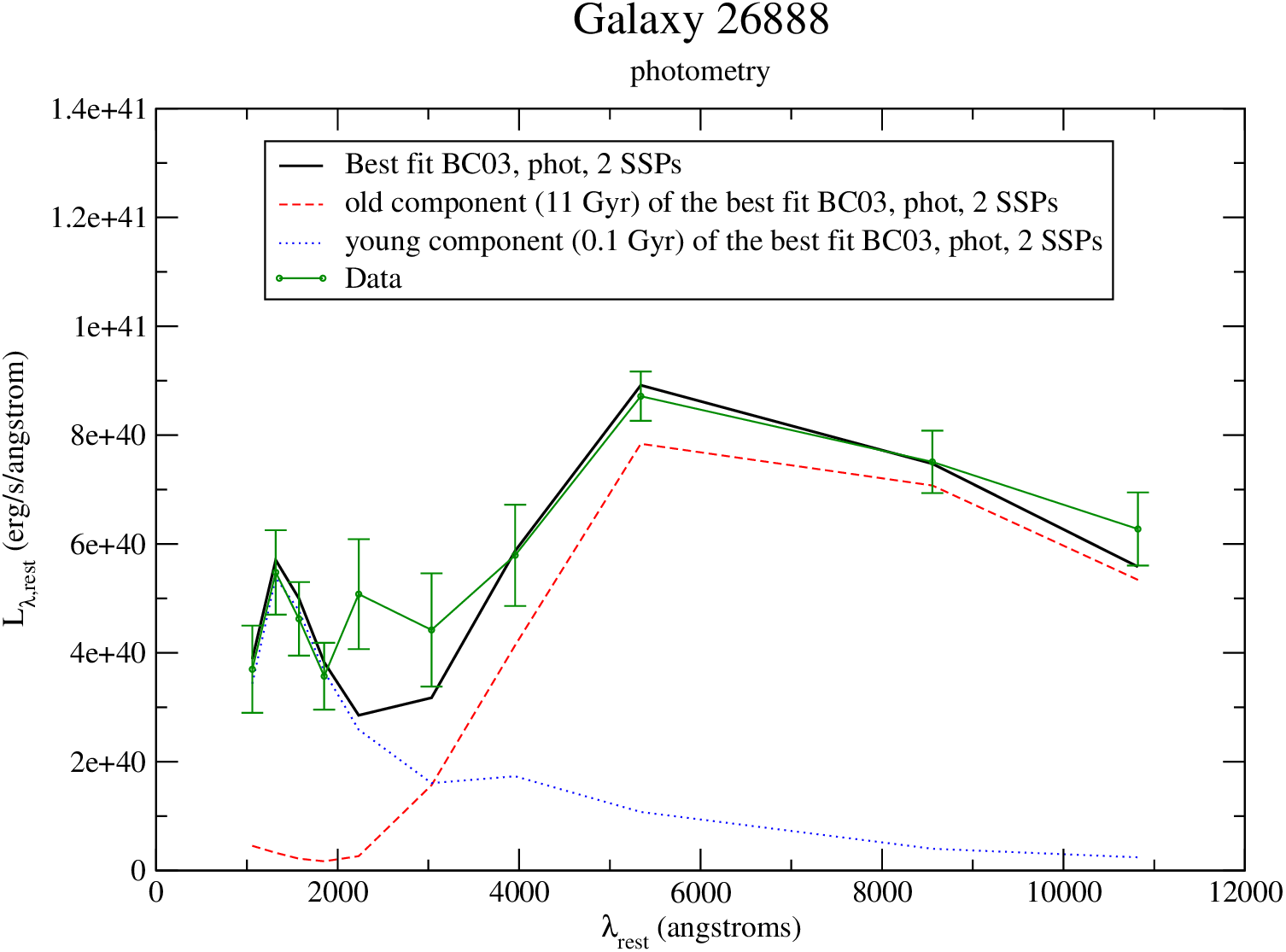}
\hspace{1cm}
\includegraphics[width=8cm]{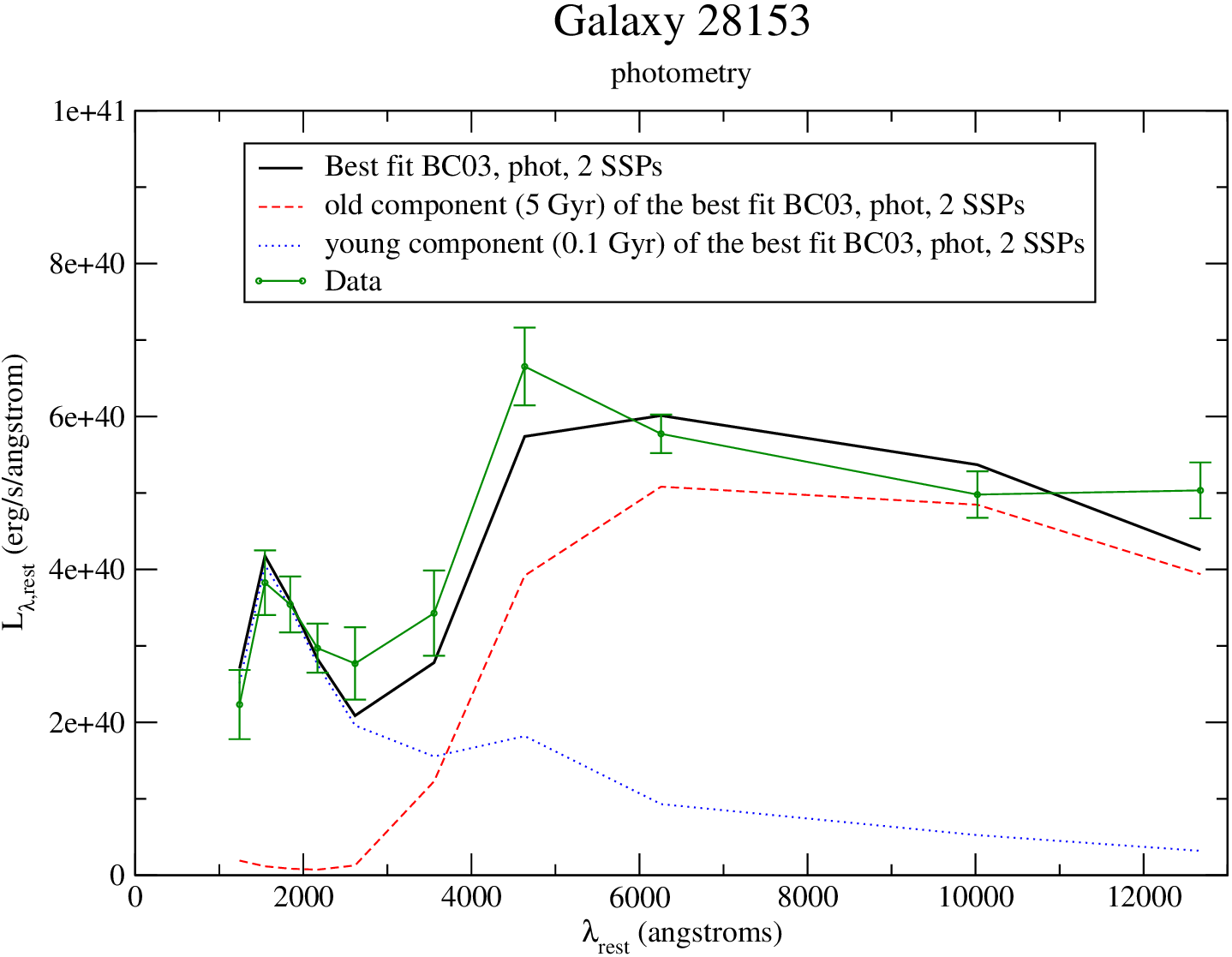}\\
\vspace{1cm}
\includegraphics[width=8cm]{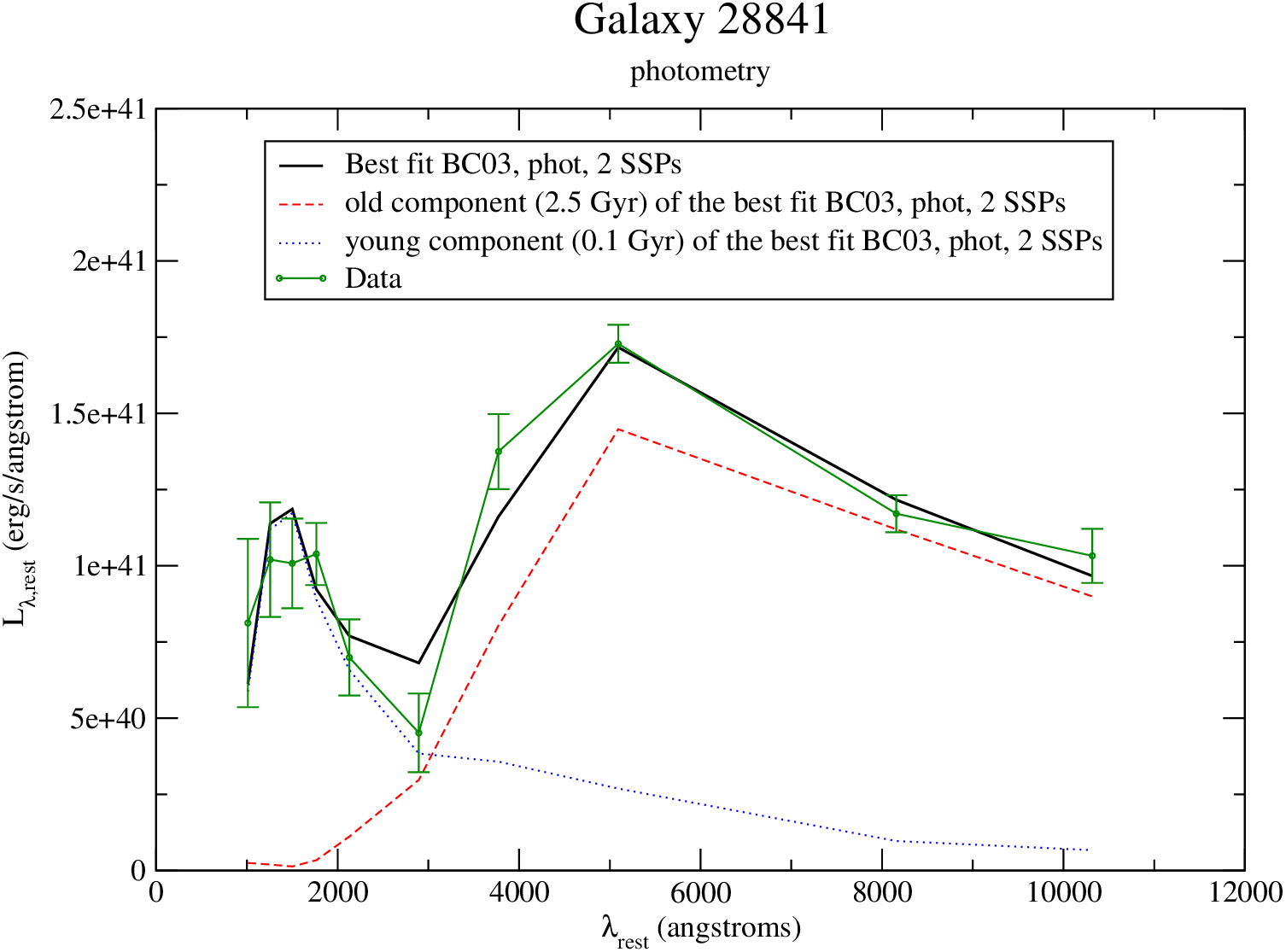}
\hspace{1cm}
\includegraphics[width=8cm]{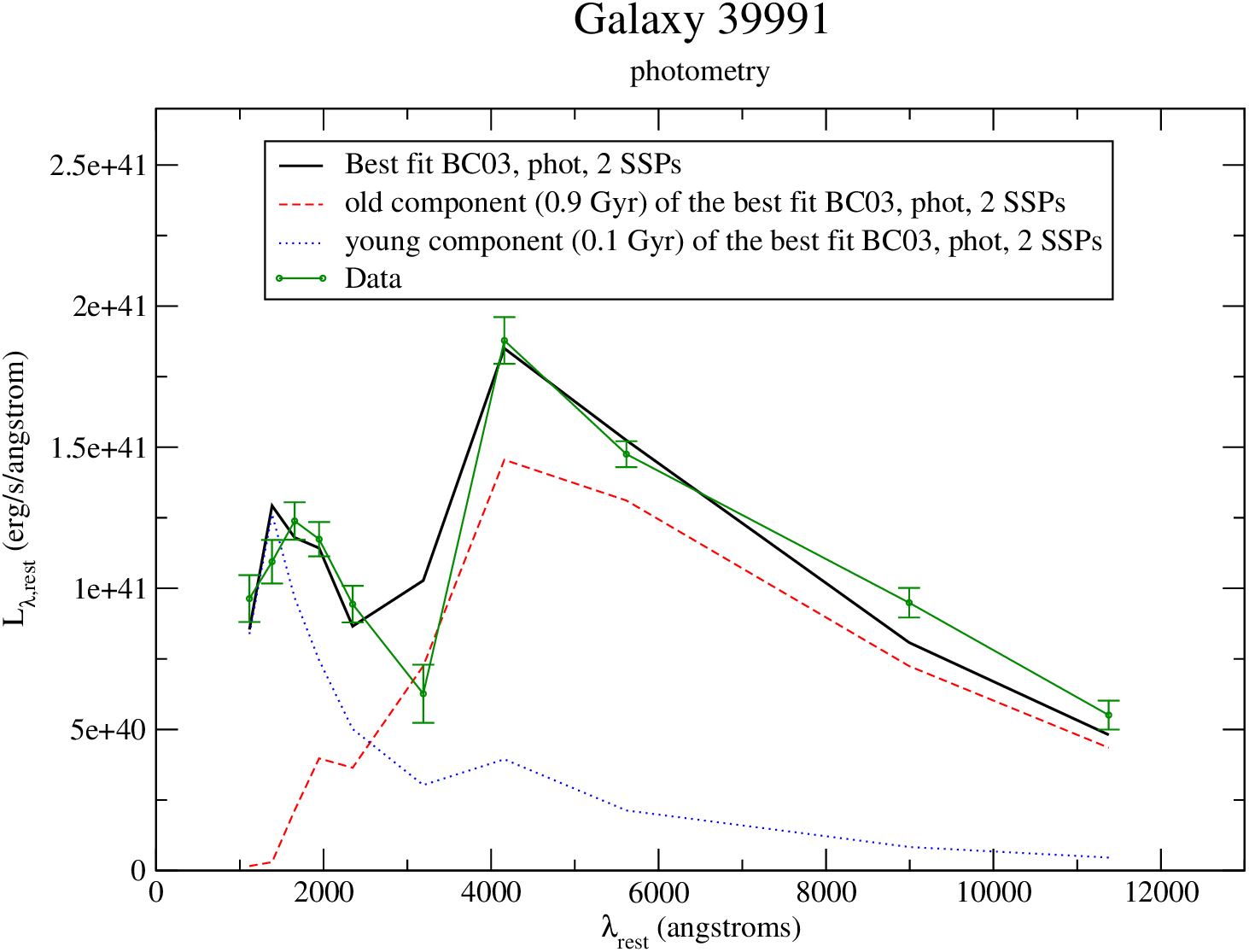}\\
Fig. \ref{Fig:2SSPlothers} (cont.)
\end{figure*}

\begin{figure*}
\vspace{1cm}
\centering
\includegraphics[width=8cm]{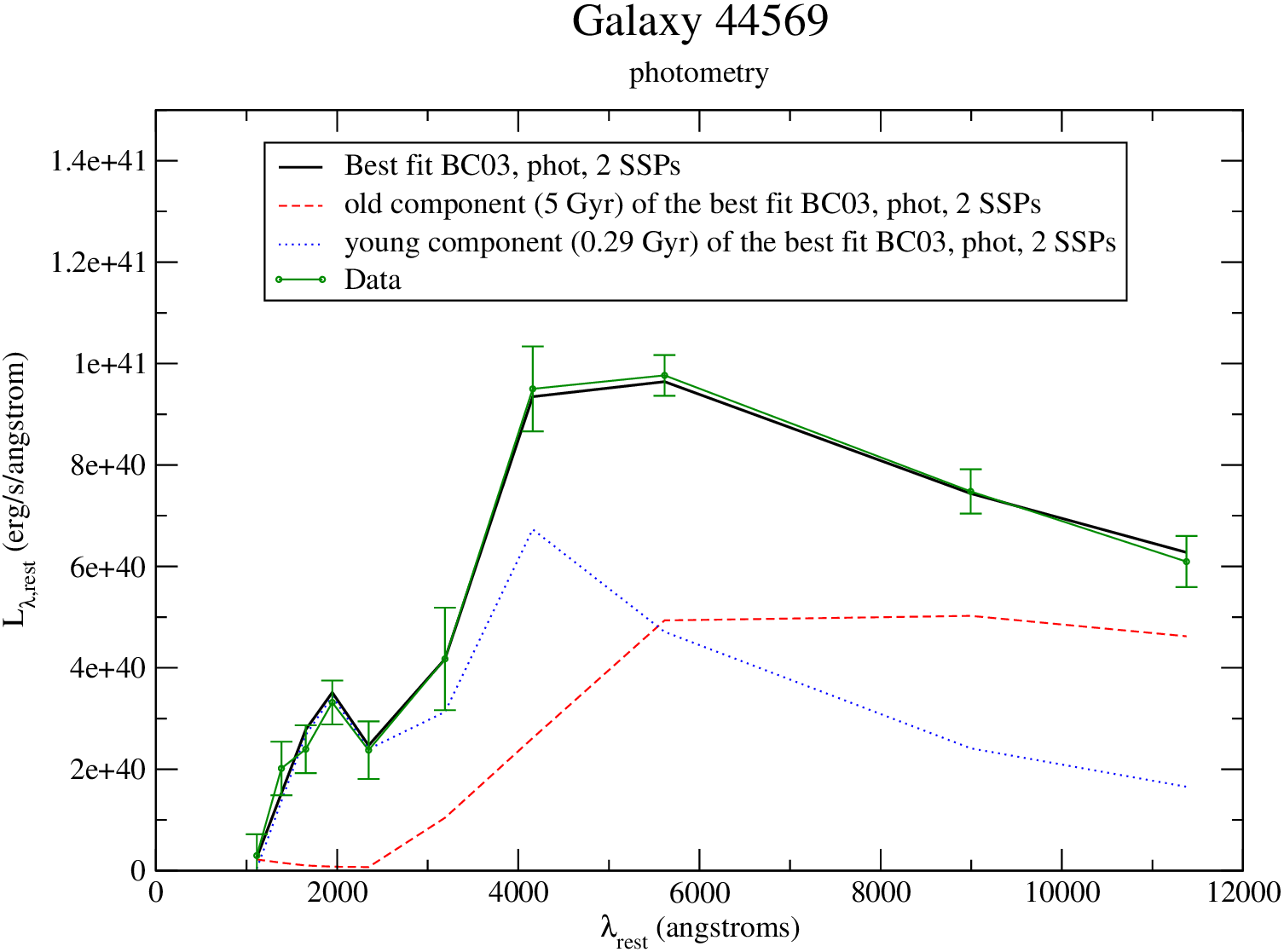}
\hspace{1cm}
\includegraphics[width=8cm]{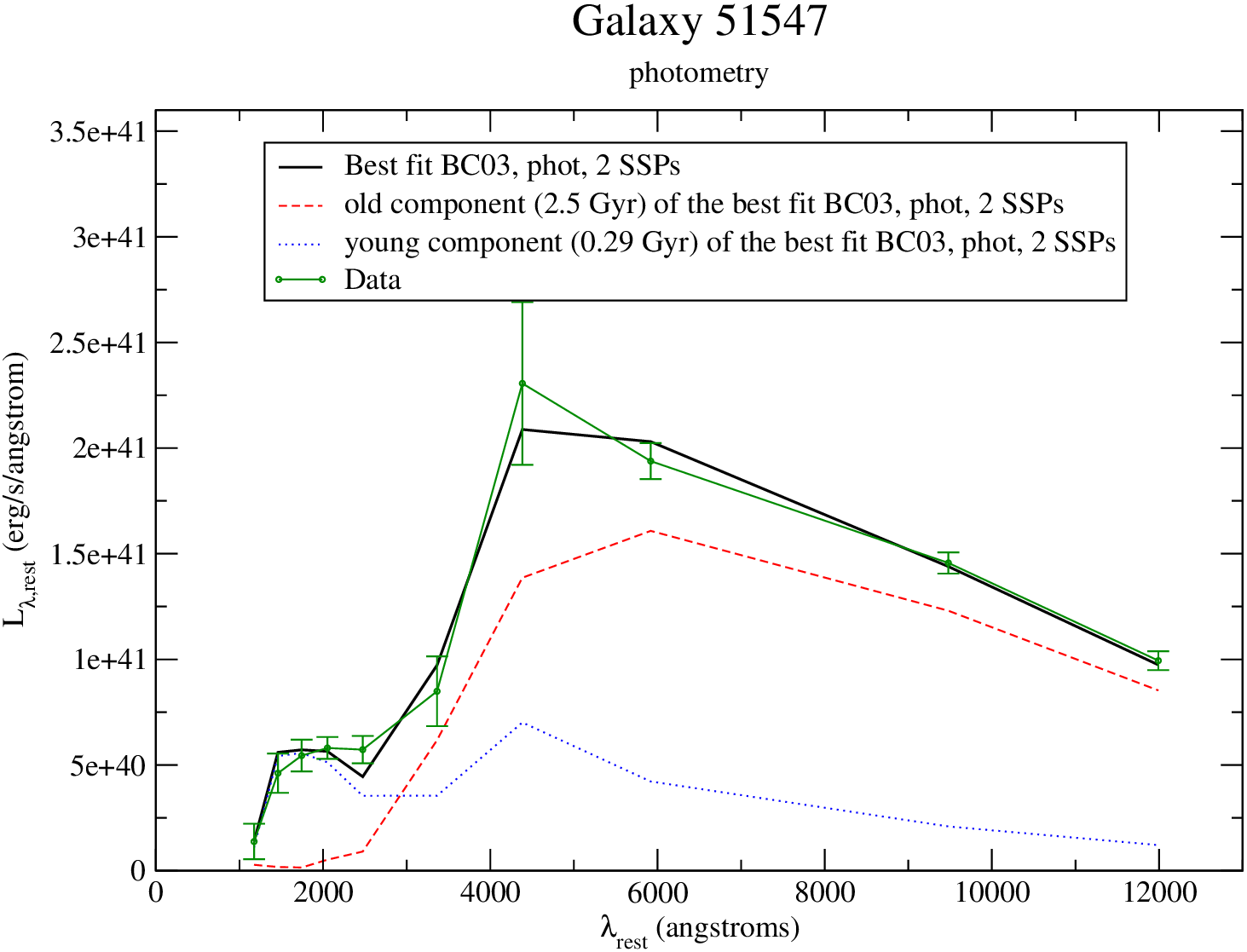}\\
\vspace{1cm}
\includegraphics[width=8cm]{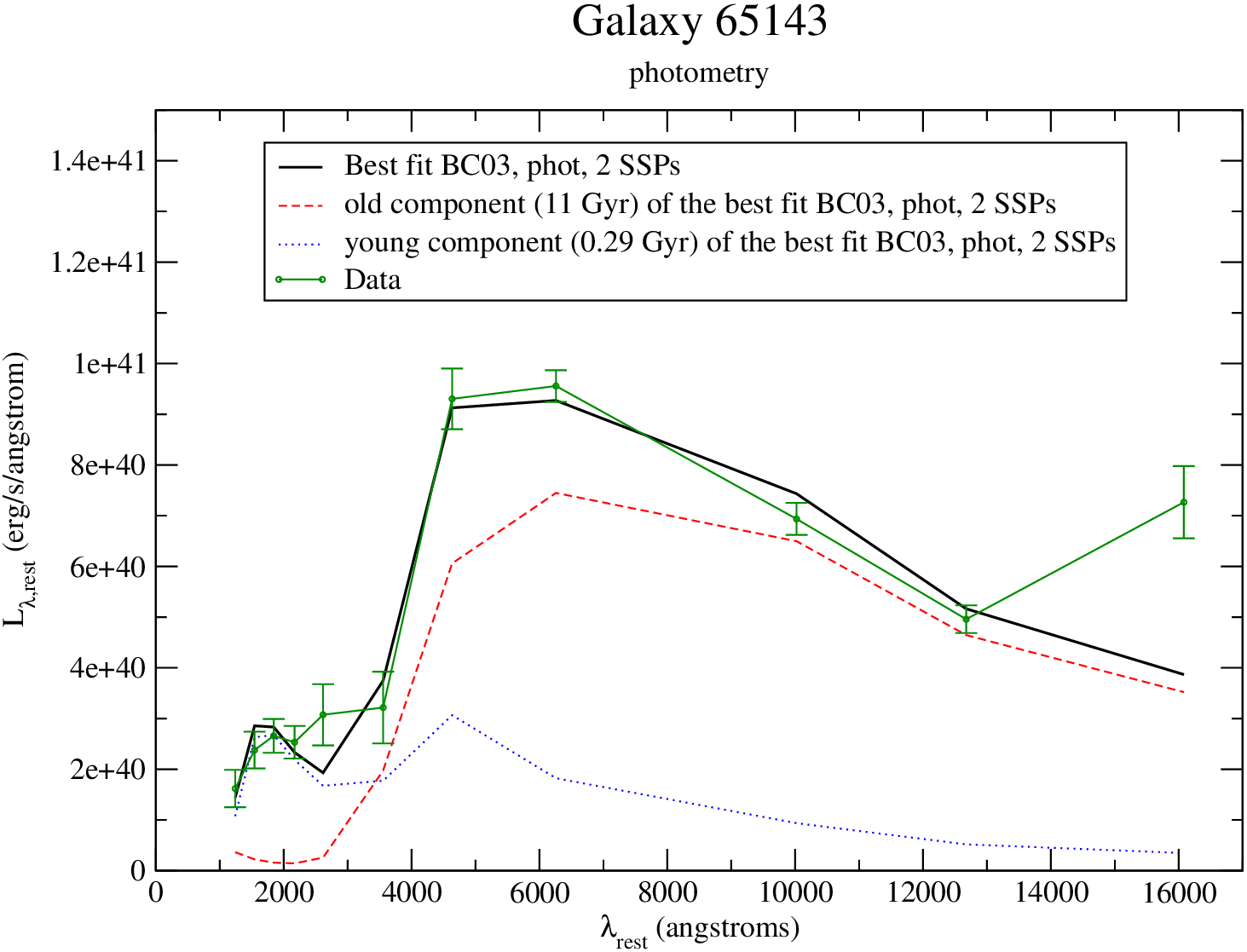}
\hspace{1cm}
\includegraphics[width=8cm]{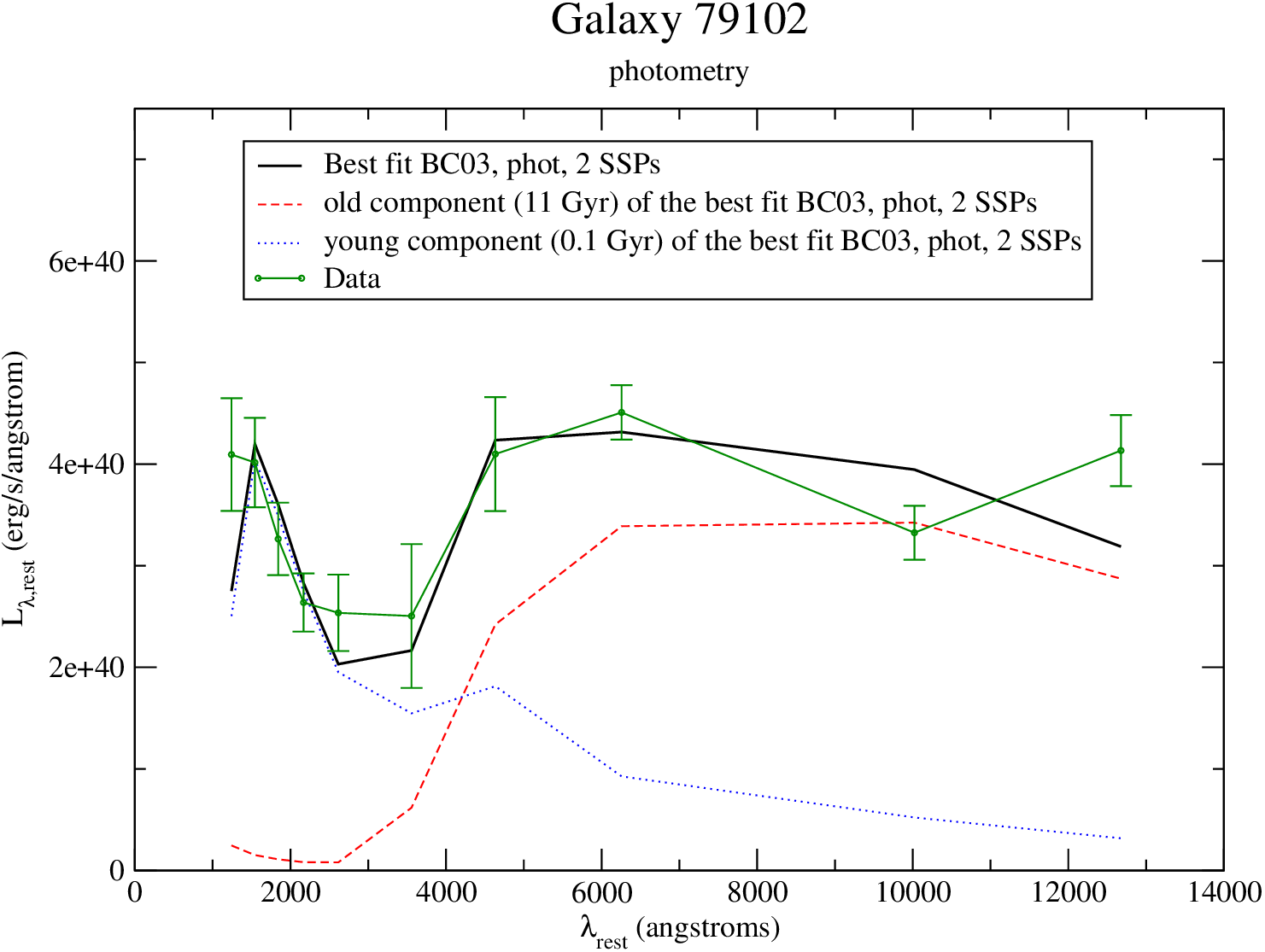}\\
\vspace{1cm}
\includegraphics[width=8cm]{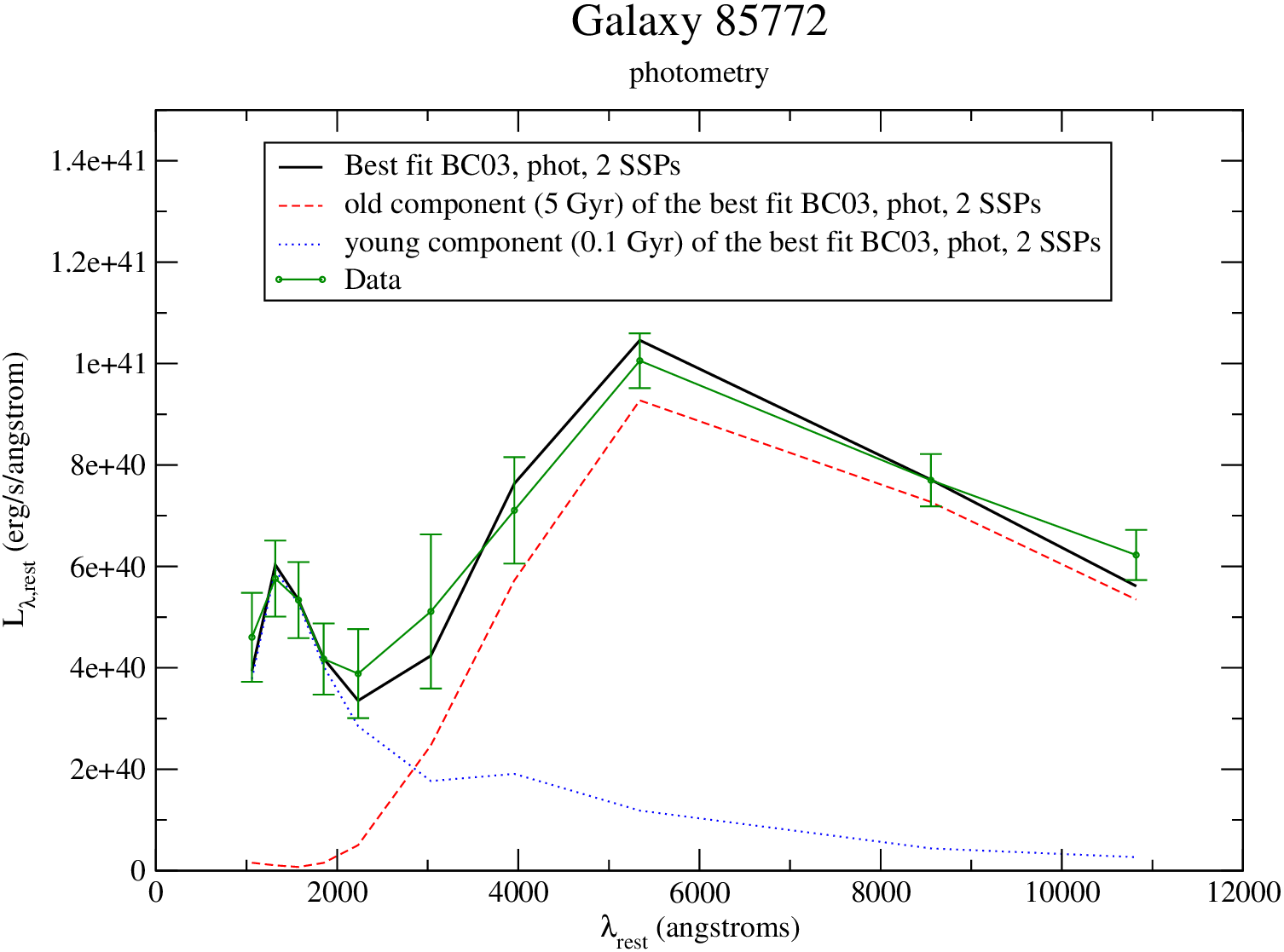}
\hspace{1cm}
\includegraphics[width=8cm]{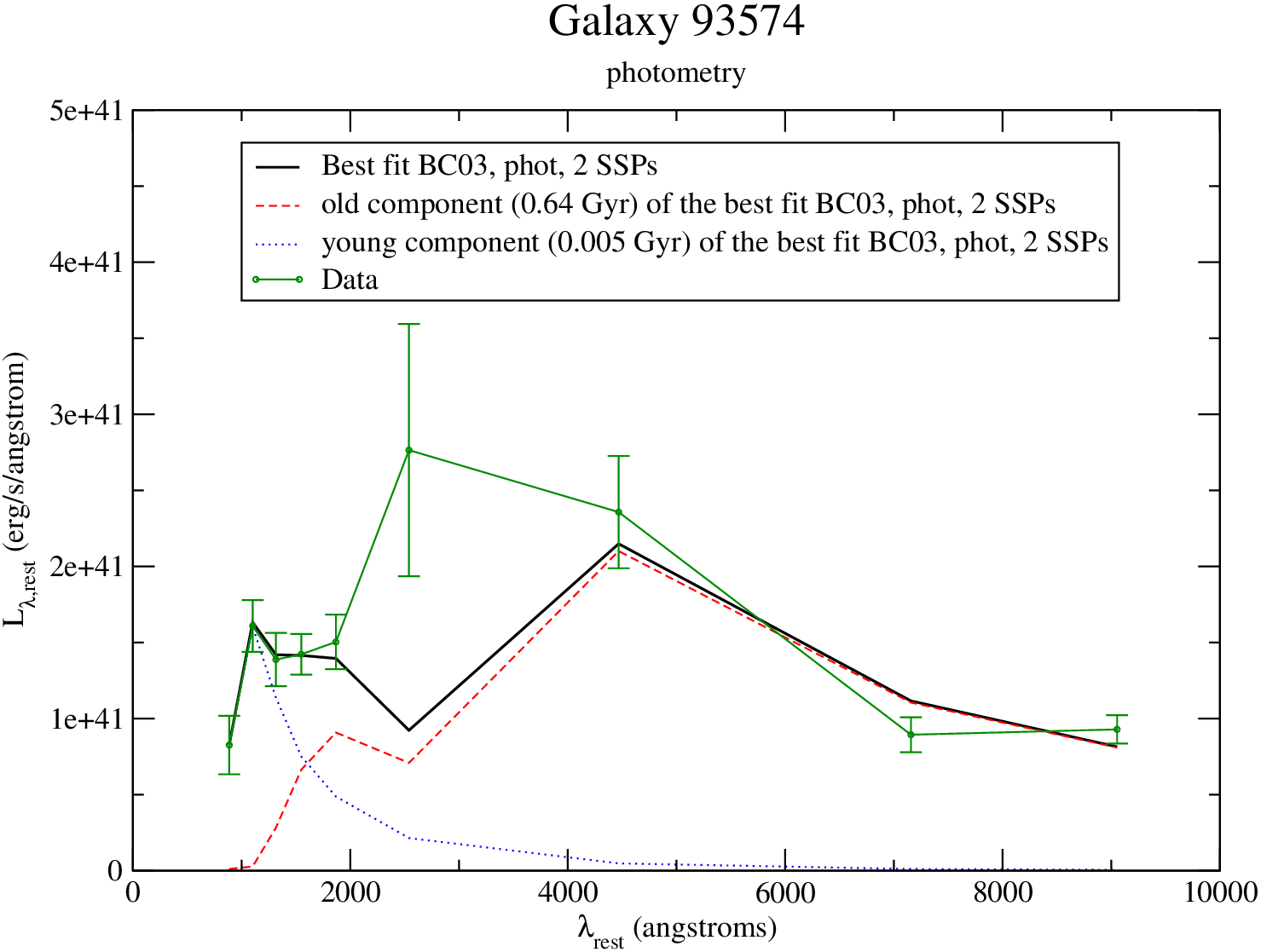}\\
Fig. \ref{Fig:2SSPlothers} (cont.)
\end{figure*}


\begin{thebibliography}{99}

\bibitem{} Alcalde Pampliega, B., P\'erez-Gonz\'alez, P. G., Dom\'\i nguez S\'anchez, H., Esquej, P., Eliche-Moral, M. C., \& Barro, G. 2015, 
in: Highlights of Spanish Astrophysics VIII, A. J. Cenarro, F. Figueras, C. Hernández-Monteagudo, J. Trujillo Bueno, L. Valdivielso, Eds., p. 360

\bibitem{} Atlee, D. W., Assef, R. J., \& Kochanek, C. S. 2009, ApJ, 694, 1539

\bibitem{} Avni, Y. 1976, ApJ, 210, 642

\bibitem{} Bedregal, A. G., Scarlata, C., Henry, A. L., et al. 2013,
ApJ, 778, id. 126

\bibitem{} Belli, S., Newman, A. B., \& Ellis, R. S. 2015, ApJ, 799, 206

\bibitem{} Bertelli, G., Bressan, A., Chiosi, C., Fagotto, F., \& 
Nasi, E., 1994, A\&AS 106, 275

\bibitem{} Bruzual, G. 2007, in: Stellar Populations as Building Blocks of Galaxies,
ed. A. Vazdekis, \& R. Peletier, Cambridge Univ. Press, Cambridge, p.\ 125

\bibitem{} Bruzual, G., \& Charlot, S. 2003, MNRAS, 344, 1000 (BC03)

\bibitem{} Calzetti, D., Armus, L., Bohlin, R. C., Kinney, A. L., Koornneef, J., 
\& Storchi-Bergmann, T. 2000, ApJ, 533, 682

\bibitem{} Castro-Rodr\'\i guez, N., L\'opez-Corredoira, M. 2012, A\&A, 537, A31 (CL12)

\bibitem{} Chabrier, G. 2003, PASP, 115, 763

\bibitem{} Cimatti, A., Cassata, P., Pozzetti, L., et al. 2008, A\&A, 482, 21  

\bibitem{} Cimatti, A., Daddi, E., Renzini, A., et al. 2004, Nature, 430, 184

\bibitem{} da Cunha, E., Charlot, S., \& Elbaz, D. 2008, MNRAS, 388, 1595
\bibitem{} Dom\'\i nguez S\'anchez, H., P\'erez-Gonz\'alez, P. G., Esquej, P., et al.
2016, MNRAS, 457, 3743

\bibitem{} Fang, G.-W., Kong, X., \& Wang, M. 2009, RAA, 9, 59

\bibitem{} Geier, S., Richard, J., Man, A. W. S, Kr\"uhler, T.,
Toft, S., Marchesini, D., \& Fynbo, J. P. U. 2013, ApJ, 777, 87

\bibitem{} Guo, Q., et al. 2011, MNRAS, 413, 101

\bibitem{} Gregg, M. D., Silva, D., Rayner, J., et al. 2006,
in The 2005 HST calibration workshop: Hubble after the transition to two-gyro mode (NASA/CP2006-214134), A. M. Koekemoer, P. Goudfrooij \& L. L. Dressel, eds., National Aeronautics and Space Administration, Goddard Space Flight Center, p.\ 209

\bibitem{} Kaviraj, S., Schawinski, K., Devriendt, J. E. G., et al.
2007, ApJS, 173, 619

\bibitem{} Koleva, M., \& Vazdekis, A. 2012, A\&A, 538, A143

\bibitem{} La Barbera, F., Ferreras, I., Vazdekis, A., de la Rosa, I. G., de Carvalho, R. R., Trevisan, M., Falc\'on-Barroso, J., \& Ricciardelli, E. 2013, MNRAS 433, 3017

\bibitem{Lab05} Labb\'e, I., Huang, J., Franx, M., et al. 2005, ApJ,
624, L81

\bibitem{} Le Borgne, J.-F., Bruzual, G., Pell\'o, R., et al.
2003, A\&A, 402, 433 

\bibitem{Lon05} Longhetti, M., Saracco, P., Severgnini, P., et al. 2005,
MNRAS, 361, 897

\bibitem{} Lonoce, I., Longhetti, M., Maraston, C., et al. 2015, MNRAS, 454, 3912

\bibitem{} L\'opez-Corredoira, M. 2010, AJ, 139, 540

\bibitem{} Maraston, C. 2005, MNRAS, 362, 799

\bibitem{} McCarthy, P. J., Le Borgne, D., Crampton, D., et al. 2004, ApJ, 614, L9

\bibitem{} Mendel, J. T., Saglia, R. P., Bender, R., et al.
2015, ApJ, 804, L4

\bibitem{} Mignoli, M., Cimatti, A., Zamorani, G., et al. 2005, A\&A, 437, 883

\bibitem{} Moresco, M., Cimatti, A., Jim\'enez, R., et al. 2012, JCAP, 8, id. 6

\bibitem{} Moresco, M., Pozzetti, L., Cimatti, A., et al. 2016, JCAP, 5, id. 14

\bibitem{} Oke, J. B. 1990, AJ, 99, 1621

\bibitem{} Onodera, M., Carollo, C. M., Renzini, A., et al. 2015, 
ApJ, 808, 161

\bibitem{} Oser, L., Ostriker, J. P., Naab, T., Johansson, P. H., \& 
Burkert, A. 2010, ApJ, 725, 2312

\bibitem{} P\'erez-Gonz\'alez, P. G., Rieke, G. H., Villar, V., et al.
2008, ApJ, 675, 234

\bibitem{} Pozzetti, L., \& Mannucci, F. 2000, MNRAS, 317, L17

\bibitem{} Renzini, A. 2006, ARA\&A, 44, 141

\bibitem{Rod07} Rodighiero, G., Cimatti, A., Franceschini, A.,
Brusa, M., Fritz, J., \& Bolzonella, M. 2007,
A\&A, 470, 21

\bibitem{} R\"ock, B., Vazdekis, A., Ricciardelli, E., Peletier, R. F., Knapen, J. H.,
\& Falc\'on-Barroso, J. 2016, A\&A, 589, A73

\bibitem{} Rowan-Robinson, M., Babbedge, T., Oliver, S., et al. 2008, MNRAS, 386, 697

\bibitem{} Rowan-Robinson, M., Gonz\'alez-Solares, E., Vaccari, M., \& Marchetti, L. 2013,
MNRAS, 428, 1958

\bibitem{} Schiavon, R. P., Faber, S. M., Rose, J. A. \& Castilho, B. V.
2002, ApJ, 580, 873

\bibitem{} Schlafly, E. F., \& Finkbeiner, D. P. 2011, ApJ, 737, id. 103

\bibitem{} Simon, J. Verde, L., \& Jim\'enez, R. 2005, Phys. Rev. D, 71, 123001

\bibitem{Ste16} Steinhardt, C. L., Capak, P., Masters, D., \& Speagle, J. S. 2016
ApJ, 824, id. 21

\bibitem{} Taylor, E. N., Franx, M., van Dokkum, P. G., et al. 2009, ApJS, 183, 295

\bibitem{} Tinsley, B. M. 1980, 
Fundamentals of Cosmic Physics, 5, 287

\bibitem{Tof05} Toft, S., van Dokkum, P., Franx, M., Thompson, R.I., 
Illingworth, G.D., Bouwens, R.J., \& Kriek, M. 2005, ApJ,
624, L9

\bibitem{} Toft, S., Gallazzi, A., Zirm, A., Wold, M., Zibetti, S., Grillo, C., \& Man, A. 2012, ApJ, 754, 3

\bibitem{} Trayford, J. W., Theuns, T., Bower, R. G., et al.
2015, MNRAS, 452, 2879

\bibitem{} Tremonti, C. 2003, PhD thesis, Johns Hopkins Univ.
 
\bibitem{Tru06} Trujillo, I., Feulner, G., Goranova, Y., et al. 2006,
MNRAS, 373, L36

\bibitem{} van de Sande, J., Kriek, M., Franx, M., et al. 2013,
ApJ, 771, id. 85

\bibitem{} Vazdekis, A., Casuso, E., Peletier, R. F., \& Beckman, J. E. 1996, ApJS, 106, 307


\bibitem{} Vazdekis, A., Peletier, R. F., Beckman, J. E., \& Casuso, E. 1997, ApJS, 111, 203

\bibitem{} Vazdekis, A., Salaris, M., Arimoto, N. \& Rose, J. A. 2001,
ApJ, 549, 274

\bibitem{} Vazdekis, A., Ricciardelli, E., Cenarro, A. J., Rivero-Gonz\'alez, J. G., 
D\'\i az-Garc\'\i a, L. A., \& Falc\'on-Barroso, J. 2012, MNRAS, 424, 157
\bibitem{} Vazdekis, A., Koleva, M., Ricciardelli, E.,
R\"ock, B., \& Falc\'on-Barroso, J. 2016, MNRAS, 463, 3409

\bibitem{} Whitaker, K. E., van Dokkum, P. G., Brammer, G., et al.
2013, ApJ, 770, L39

\bibitem{Wik08} Wiklind, T., Dickinson, M., Ferguson, H.C.,
Giavalisco, M., Mobasher, B., Grogin, N.A., \& Panagia, N. 2008,
ApJ, 686, 781

\bibitem{} Williams, R. J., Quadri, R. F., Franx, M., van Dokkum, P., \& Labb\'e, I. 2009, ApJ, 691, 1879

\bibitem{} Yi, S. K., Lee, J., Sheen, Y.-K., Jeong, H., Suh, H., \& Oh, K.
2011, ApJS, 195, 22

\end{thebibliography}
\end{document}